\documentclass[]{JHEP}
\usepackage{epsfig}

\def\be{\begin{equation}}
\def\ee{\end{equation}}
\def\bea{\begin{array}}
\def\eea{\end{array}}
\def\beqa{\begin{eqnarray}}
\def\eeqa{\end{eqnarray}}
\def\beqas{\begin{eqnarray*}}
\def\eeqas{\end{eqnarray*}}

\def\bp{\begin{picture}}
\def\ep{\end{picture}}
\def\bc{\begin{center}}
\def\ec{\end{center}}
\def\bfig{\begin{figure}}
\def\efig{\end{figure}}

\def\bit{\begin{itemize}}
\def\eit{\end{itemize}}

\def\f{\frac}
\def\sq{\sqrt}

\def\[{\left[}
\def\]{\right]}
\def\({\left(}
\def\){\right)}

\def\..{\left.}
\def\.{\right.}
\def\ra{\rightarrow}
\def\la{\leftarrow}

\def\tm{\times}

\def\da{\dagger}

\def\la{\lambda}

\def\al{\alpha}
\def\bt{\beta}

\def\ep{\epsilon}

\def\pa{\partial}
\def\pr{\prime}
\def\eqv{\equiv}

\title{The $SU(3)_C{\times} SU(4)_W {\times} U(1)_{B-L}$
Models with Left-Right Unification }

\author{Tianjun Li$^{1,2}$, Fei Wang$^3$, Jin Min Yang$^1$ \\
 $^1$ Key Laboratory of Frontiers in Theoretical Physics, \\
      ~~ Institute of Theoretical Physics, Academia Sinica, Beijing 100190, China \\
$^2$ George P. and Cynthia W. Mitchell Institute for Fundamental Physics, \\
     ~~ Texas A$\&$M University, College Station, TX 77843, USA\\
$^3$ School of Physics, Monash University, Melbourne 3800, Australia}

\received{\today}
\accepted{\today}

\preprint{arXiv:0901.2161}

\abstract{ To understand the origin of the left-right symmetry, we
study a partial unification model based on
$SU(4)_W{\times}U(1)_{B-L}$ which can be broken down to the minimal
left-right model either through the Higgs mechanism in four
dimensions or through the five-dimensional orbifolding gauge
symmetry breaking,especially we propose to use the rank reducing
outer automorphism orbifolding breaking mechanism.
 We scrutinized all these breaking
mechanisms and found that for the orbifold breaking in five
dimensions, the rank-reducing outer automorphism is better than the
inner automorphism and can make the low energy theory free of the
$U(1)_Z$ anomaly.It is possible for the outer automorphism
orbifolding breaking mechanism to be non-anomalous without
Chern-Simons terms and new localized fermions. For the
four-dimensional model with the Higgs mechanism, we study in detail
both its structure and its typical phenomenology. It turns out that
this four-dimensional scenario may predict some new phenomenology
since the new mirror fermions (which are introduced in order to fill
the SM fermions into $SU(4)_W$ without anomaly) are preserved at low
energy scale and mix with the SM fermions. We also examine the gauge
coupling unification in each case, and discuss the possibility for
unifying this partial unification group with the Pati-Salam group
$SU(4)_{PS}$ to realize a grand unification.}


\begin{document}
\maketitle

\newpage
\section{Introduction}
Although the Standard Model (SM) of the electroweak interaction
based on the spontaneously broken gauge symmetry
$SU(2)_L{\times}U(1)_Y$ has been extremely successful in describing
the phenomena below weak scale, it leaves many theoretical or
aesthetical questions unanswered, one of which is the origin of
parity violation. We want to know the reason why the weak
interaction apparently violates parity while all other forces
conserve parity, and whether parity conservation can be achieved at
a more fundamental level. On the other hand, the discovery of
neutrino masses through neutrino oscillation experiments also
requires an explanation beyond the SM. Both questions can be
elegantly addressed in the so-called left-right models which restore
the left-right symmetry at some high energy scale and broken down to
the SM at the weak scale.

Among the left-right models the minimal left-right model based on
$SU(2)_L \times SU(2)_R \times U(1)_{B-L}$ is
most popular and has been extensively studied \cite{mohapatra}. The
key assumption in this model is that the fundamental weak
interaction is invariant under parity symmetry and the
observed parity violation is the consequence of the spontaneous
breaking of parity symmetry. Such a hypothesis requires the existence
of right-handed neutrinos and thus can give massive neutrinos.
However, in this model
the parity invariance has to be put in by hand and there is no ad hoc reason
why $SU(2)_L$ coupling should be identical to $SU(2)_R$ coupling.
Only in some Grand Unification Theories (GUT) like the $SO(10)$ models \cite{so10}
can the equality of the two $SU(2)$ gauge couplings be naturally
guaranteed through the gauge coupling unification at a much
higher energy scale.

Of course, from the phenomenological point of view we do not know
in prior what really happens in high energy region.
So the partially unified models like the Pati-Salam model based on
$SU(4)_C{\times}SU(2)_L{\times}SU(2)_R$ \cite{ps}
are quite interesting in the sense
that they can provide a bridge between low energy theory and high energy
GUT theory.
Note that in the Pati-Salam partial unification model the parity invariance
is also put in by hand.
In this work we study an alternative model based on
$SU(4)_W \times U(1)_{B-L} \times SU(3)_C$ for
the partial unification of left-right gauge groups.
The partially unified models, where $SU(2)_L\times SU(2)_R$
is unified into a semi-simple group, can give an explanation
for the origin of parity symmetry. Note that the $SO(4)$
is not a simple group, the minimal $SU(2)_L\times SU(2)_R$
unification models are $SO(5)$ and have been studied
previously~\cite{Agashe:2004rs}.
In this paper, we consider the scenarios where
 the $SU(2)_L$ and $SU(2)_R$ gauge groups are
embedded into $SU(4)_W$. And the exact left-right symmetry is
naturally realized in the $SU(4)_W$ invariant Lagrangian. When the
gauge groups are one-step further unified into
$SU(4)_W{\times}SU(4)_{PS}$, such a partial unification can lead to
the typical rank-six simple group unification (for example, $SO(12)$
gauge group) at higher energy scales.

Such a partial unification idea for left-right symmetry was first
proposed in five dimensions \cite{shafi,nandi}. In our study we
focus on $SU(4)_W \times U(1)_{B-L} \times SU(3)_C$ and discuss
various possibilities for symmetry breaking, including the Higgs
mechanism in four dimensions and orbifolding gauge symmetry breaking
in five dimensions \cite{orbifold1,orbifold2,hebecker,hebecker2,
Li:2001qs, Li:2001wz},especially we propose to use rank reducing
outer automorphism orbifolding breaking which is not tried in the
literature.We find that for the orbifold breaking in five
dimensions, the outer automorphism can have the advantage to keep
the low energy theory free from the anomalous $U(1)_Z$.Thus it is
possible for the outer automorphism orbifolding breaking mechanism
to be non-anomalous without Chern-Simons terms and new localized
fermions. For the four-dimensional model with the Higgs mechanism,
new mirror fermions have to be introduced in order to fill the SM
fermions into $SU(4)_W$ without anomaly. These mirror fermions are
preserved at low energy scale and their mixings with the SM fermions
may have rich phenomenological consequences.

The content of this work is organized as follows. In Sec.
\ref{sec-2}, we discuss the $SU(4)_W$ left-right unification model,
focusing on the gauge symmetry breaking mechanisms through the
five-dimensional  orbifold gauge symmetry
 breaking mechanism,especially we propose to use the rank reducing outer automorphism orbifolding breaking
in addition to the mostly used inner automorphism breaking
mechanism;and also we study in detail the Higgs mechanism in four
dimensions. In Sec. \ref{sec-3}, we examine the running and the
unification of the gauge couplings in each case, and discuss the
possibility for unifying this partial unification group with the
Pati-Salam group $SU(4)_{PS}$ to realize a grand unification In Sec.
\ref{sec-4}, we briefly discuss the phenomenology of the
four-dimensional theory. Sec.  \ref{sec-5} is our conclusion.

\section{$SU(4)_W{\times}U(1)_{B-L}$ Left-Right Unification Model }
\label{sec-2}
\subsection {Basic structure of the model}
In the minimal left-right model based on $SU(2)_L{\times}SU(2)_R{\times}U(1)_{B-L}$,
the left-handed and right-handed fermions have $SU(2)_L$ and $SU(2)_R$
gauge interactions, respectively. When the two gauge groups are
unified into $SU(4)_W$, the matter content has to be embedded into
some representations of $SU(4)_W$. As the gauge group commutate with
the Lorentz group, only one type of chiral states (left or right)
is allowed in one gauge multiplet (note that the matter content is not
filled properly in \cite{bern} so that the gauge group does not
commute with Lorentz group).  Besides, the baryon number conservation
at low energy requires the existence of new fermions in the
representations in addition to the SM fermions.
We introduce fundamental (and anti-fundamental) representation of $SU(4)_W$ as
\begin{eqnarray}
&& ({\bf 4}): X_f \sim \left( \begin{array}{c}
  u_L\\ d_L\\ U_L^c\\ -D_L^c \end{array} \right)_{\frac{1}{3}}
~~;~~({\bf 4^*}):Y_f \sim \left(
  \begin{array}{c}
  U_L\\ -D_L\\ u_L^c\\ d_L^c \end{array} \right)_{-\frac{1}{3}}~~;~ \nonumber \\
&&({\bf 4}): L_f \sim \left( \begin{array}{c}
  {\nu}_L\\ e_L\\ N^c_L\\ -E_L^c \end{array}\right)_{-1}
~~;~~({\bf 4^*}): N_f \sim \left( \begin{array}{c}
   N_L\\ -E_L\\ {\nu}_L^c\\ e_L^c \end{array} \right)_{1}~~,~\,
\end{eqnarray}
where $\phi_L^c{\equiv}(\phi^c)_L$, and the minus sign conforms to
our choice of $Q^a=(D_L, U_L)$ in $SU(2)_L$ representations ${\bf
2}$ and being related to its conjugate by $Q_a=(U_L,-D_L)$ through
antisymmetric tensor $Q^a=\epsilon^{ab}Q_b$.

The $SU(2)_L{\times}SU(2)_R$ components in $SU(4)_W$ can be seen
by decomposing the representation into
\begin{eqnarray}
\bf{4}=(\bf{2,1}){\oplus}(\bf{1,2})~.~\,
\end{eqnarray}
The representation of each fermion in
$SU(3)_C{\times}SU(2)_L{\times}SU(2)_R{\times} U(1)_{B-L}$ is
listed in Table \ref{rep}. The electric charge is an additive quantum number
and is a linear combination of the diagonal generators.
From the representation of the fermions we can get the formula for the
electric charge
\begin{eqnarray}
Q=T_{3L}+T_{3R}+\frac{1}{2}Y_{B-L}~.~\,
\end{eqnarray}

\begin{table}
\caption{Representations of fermion fields under gauge groups
$SU(3)_C,SU(2)_L,SU(2)_R$, $U(1)_{B-L}$ and $U(1)_Q$.}
\begin{center}
\begin{tabular}{|c|c|c|c|c|c|}
\hline ~~~~~~~&$SU(3)_C$&$SU(2)_L$&$SU(2)_R$&$U(1)_{B-L}$&$Q$\\
\hline $(\nu_L~~e_L)_f$&\bf{1}&\bf{2}&\bf{1}& $-1$&$(0~~-1)$\\
\hline $(N_L~~-E_L)_f$&\bf{1}&\bf{2}&\bf{1}&1&$(0~~1)$\\ \hline
$(e^c_L~~-\nu^c_L)_f$&\bf{1}&\bf{1}&$\bf{2}$&1&(1~~0)\\ \hline
$(N^c_L~~-E_L^c)_f$&\bf{1}&\bf{1}&$\bf{2}$&$-1$&$(0~~-1)$\\
\hline $(u_L~~d_L)_f$&\bf{3}&\bf{2}&$\bf{1}$&
$\frac{1}{3}$&$(\frac{2}{3}~-\frac{1}{3})$\\ \hline
$(U_L~~-D_L)_f$&$\bf{3}^*$&\bf{2}&\bf{1}&$-\frac{1}{3}$&$(-\frac{2}{3}~~\frac{1}{3})$\\
\hline
$(d^c_L~~-u^c_L)_f$&$\bf{3}^*$&\bf{1}&$\bf{2}$&$-\frac{1}{3}$&
  $(\frac{1}{3}~-\frac{2}{3})$\\ \hline
$(U_L^c~~-D_L^c)_f$&$\bf{3}$&\bf{1}&$\bf{2}$&$\frac{1}{3}$&
 $(\frac{2}{3}~-\frac{1}{3})$\\ \hline
\end{tabular}
\end{center}
\label{rep}
\end{table}

The normalization of the generators in the fundamental representation reads
\begin{eqnarray}
Tr(T^aT^b)=\frac{1}{2}\delta^{ab} \, .
\end{eqnarray}
We can choose the $SU(4)_W$ generators in the form
\begin{eqnarray}
T_{3L}=\f{1}{2}\left( \begin{array}{cccc}
  1&&&\\ &-1&&\\ &&0&\\ &&&0 \end{array} \right),
~~T_{3R}=\f{1}{2}\left( \begin{array}{cccc}
  0&&&\\ &0&&\\ &&1&\\ &&&-1 \end{array} \right),
~~ T_Z=\f{\sqrt{2}}{4}\left( \begin{array}{cccc}
  1&&&\\ &1&&\\ &&-1&\\ &&&-1 \end{array} \right).
\end{eqnarray}
The $U(1)_Z$ charge assignment of the fundamental representation
can be written as
\begin{eqnarray}
 Y_Z=\left( \begin{array}{cccc}
   1&&&\\ &1&&\\ &&-1&\\ &&&-1 \end{array} \right)~,~\,
\end{eqnarray}
and the normalization of gauge group $U(1)_Z$ reads
\begin{eqnarray}
T_{Z}=\frac{\sqrt{2}}{2}\f{Y_{Z}}{2} ~.~\,
\end{eqnarray}
So the relation between the gauge coupling $U(1)_Z$ and $SU(4)_W$ is
\begin{eqnarray}
g_{Z}=\frac{\sqrt{2}}{2}g_4 \, .
\end{eqnarray}
Since the adjoint representation of $SU(4)_W$ can be decomposed as
\begin{equation}
\bf{15} = (\bf{3,1})\oplus (\bf{1,3}) \oplus (\bf{2,\bar{2}}) \oplus
(\bf{\bar{2},2}) \oplus (\bf{1,1}) \, ,
\end{equation}
we can write the gauge part of the Lagrangian into its gauge components
($M,N$ indicate space-time index in arbitrary dimension)
\begin{eqnarray}
{\cal L} &=&-\frac{1}{4g^2}F_{M N}^aF_{MN}^a \nonumber \\
   &\sim& -\frac{1}{4g^2}F_{MN}^{iL}F^{iL}_{MN}
   -\frac{1}{4g^2}F_{MN}^{iR}F_{MN}^{iR}-\frac{1}{4g^2}F_{MN}^ZF_{MN}^Z ~.~\,
\end{eqnarray}

In the following we will discuss three different gauge symmetry
broken mechanisms: the five-dimensional  orbifold gauge symmetry
 breaking mechanism,especially we propose to use the rank reducing outer automorphism orbifolding breaking
in addition to the mostly used inner automorphism breaking
mechanism;
 and also we discuss the symmetry breaking via the Higgs mechanism in four dimensions.

\subsection{Five-dimensional inner automorphism orbifold breaking}
Consider the five-dimensional space-time ${\cal M}_4{\tm}
S^1/(Z_2{\tm}Z_2)$ comprising of Minkowski space ${\cal M}_4$ with
coordinates $x_{\mu}$ and the orbifold $S^1/(Z_2{\tm}Z_2)$ with
coordinate $y{\eqv}x_5$. The orbifold $S^1/(Z_2{\tm}Z_2)$ is
obtained by identification
\beqa
P:y{\sim} -y~,~~~~~~~~~P^{\pr}:y^{\pr}\sim -y^{\pr} ~,~\,
\eeqa
where
$y^{\pr}{\eqv}y+\pi R/2$. There are two inequivalent 3-branes locating at $y=0$
and $y=\pi R/2$ which are denoted as $O$ and
$O^{\pr}$, respectively. The action of discrete groups on the field space
is specified as
\beqa
 \phi(x_{\mu},y)&\sim& P_{\phi}\phi(x_{\mu},-y) ~,~\,\\
 \phi(x_{\mu},y^{\pr})&\sim& P_{\phi^{\pr}}\phi(x_{\mu},-y^{\pr}) ~,~\,
\eeqa
where $\phi(x_{\mu},y)$ denotes a vector comprising of bulk fields,
and $P_{\phi}$ and $P_{\phi^{\pr}}$ are the matrix representation
of the two $Z_2$ operator actions which have eigenvalues $\pm 1$.
In the diagonal basis these fields have the KK expansions as
 \beqa
\phi_{++}(x_{\mu},y)&=&\sum\limits_{n=0}^{+\infty}
  \sq{\f{1}{2^{\delta_{n,0}}\pi
R}}\phi_{++}^{(2n)}(x_\mu)\cos\f{2 n y}{R} \, ,\\
\phi_{+-}(x_{\mu},y)&=&\sum\limits_{n=0}^{+\infty}
 \sq{\f{1}{\pi R}} \phi_{+-}^{(2n+1)}(x_\mu)\cos\f{(2n+1) y}{R} \, ,\\
\phi_{-+}(x_{\mu},y)&=&\sum\limits_{n=0}^{+\infty}
 \sq{\f{1}{\pi R}} \phi_{-+}^{(2n+1)}(x_\mu)\sin\f{(2n+1)y}{R} \, ,\\
\phi_{--}(x_{\mu},y)&=&\sum\limits_{n=0}^{+\infty}
 \sq{\f{1}{\pi R}} \phi_{--}^{(2n+2)}(x_\mu)\sin\f{(2n+2)y}{R} \, ,
\eeqa where $n$ is an integer and the fields
$\phi_{++}^{(2n)}(x_\mu)$, $\phi_{+-}^{(2n+1)}(x_\mu)$,
$\phi_{-+}^{(2n+1)}(x_\mu)$, $\phi_{--}^{(2n+2)}(x_\mu)$
respectively acquire a mass $2n/R$, $(2n+1)/R$, $(2n+1)/R$ and
$(2n+2)/R$ upon compactification. Only $\phi_{++}(x_{\mu},y)$
possess a 4-D massless zero mode. It is easy to see that $\phi_{++}$
and $\phi_{+-}$ are non-vanishing at $y=0$ and $\phi_{++},\phi_{-+}$
are non-vanishing at $y=\pi R/2$.

We can choose the parity assignment of the fields in terms of the
fundamental representation of $SU(4)_W$:

 \beqa P=diag(+1,+1,+1,+1) ~,~~~
P^{\pr}=diag(+1,+1,-1,-1)~,~\, \eeqa with the transformation law
\beqa \label{transf}
P:~~ V^a_{\mu}(x_\mu,-y)&=&P V^a_\mu(x_{\mu},y)P ~,~\, \\
 P:~~ V^a_{5}(x_\mu,-y)&=&-P V^a_5(x_{\mu},y)P  ~,~\, \\
P^{\pr}:~~V^a_\mu(x_{\mu},-y^{\pr})&=&P^{\pr}
V^a_\mu(x_{\mu},y^\pr)P^{\pr} ~,~\, \\
P^{\pr}:~~V^a_5(x_{\mu},-y^{\pr})&=&-P^{\pr}
V^a_5(x_{\mu},y^\pr)P^{\pr} ~.~\, \eeqa From the assignment of $P$
and $P^{\pr}$ we know that the gauge symmetry $SU(4)_{w}$ is broken
by boundary conditions to $SU(2)_L{\tm}SU(2)_R{\tm}U(1)_X$ on the
boundary $O^{\pr}$ brane and remains in the bulk as well as on the
$O$ brane. There are three possibilities for the location of matter
fields:
\begin{itemize}
\item[(i)] The first possibility is to put matter in the bulk.
   To get the mass spectrum, we make the choice
\beqa
P:~~X(x_\mu,-y)&=& P X(x_\mu,y)  ~,~\,\\
    L(x_\mu,-y)&=& P Y(x_\mu,y)  ~,~\,\\
P:~~Y(x_\mu,-y)&=& P Y(x_\mu,y)  ~,~\,\\
    N(x_\mu,-y)&=& P N(x_\mu,y)  ~,~\,\\
P^\prime:~~X(x_\mu,-y^\pr)&=& P^\prime X(x_\mu,y^\pr)  ~,~\,\\
           L(x_\mu,-y^\pr)&=& P^\prime Y(x_\mu,y^\pr)  ~,~\,\\
P^\prime:~~Y(x_\mu,-y^\pr)&=& -P^\prime Y(x_\mu,y^\pr)  ~,~\, \\
           N(x_\mu,-y^\pr)&=& -P^\prime N(x_\mu,y^\pr) ~.~\,
\eeqa
 We denote the matter content of the fundamental and anti-fundamental
representations of $SU(4)_W$ as $X_{\bf 4}\sim(Q~,~\bar{Q^c})$,
$Y_{{\bf 4}^*}\sim(\bar{Q}~,~Q^c)$, $L_{\bf 4}\sim(L~,~\bar{L^c})$ and
$N_{{\bf 4}^*}\sim(\bar{L}~,~L^c)$, respectively. The parity
assignment of the matter fields are listed in Table \ref{new}.
\begin{table}
\begin{center}
\caption{Parity assignments for the matter fields.}
\vspace*{0.2cm}
\begin{tabular}{|c|c|c|} \hline
$(P,P^\pr)$&4-D matter fields& mass\\ \hline
$(+,+)$&$Q,Q^c,L,L^c$&$\f{2n}{R}$\\
\hline$(+,-)$&$\bar{Q}^c,\bar{Q},\bar{L},\bar{L}^c$&$\f{2n+1}{R}$\\ \hline
\end{tabular}
\label{new}
\end{center}
\end{table}
The matter content in the low energy effective theory is same as in
the minimal left-right model. However, due to the charge assignments
for $U(1)_Z$, the anomaly does not cancel in this theory. A possible
solution for this problem is to introduce Chern-Simmons terms which
can cancel the anomaly.

The Yukawa coupling can be included by introducing bulk Higgs fields.
From the parity assignments of the matter fields, to make the action
invariant under parity transformation, the adjoint Higgs field
$\Sigma_1$ and the symmetric Higgs field $\Sigma_2$, $\Sigma_3$
(in ${\bf \overline{10}}$ and ${\bf 10}$ respectively) must transform as
\beqa
P:~~ \Sigma_1(x_\mu,-y)&=&P \Sigma_1(x_{\mu},y)P  ~,~\, \\
P:~~ \Sigma_2(x_\mu,-y)&=&P \Sigma_2(x_{\mu},y)P  ~,~\, \\
P:~~ \Sigma_3(x_\mu,-y)&=&P \Sigma_3(x_{\mu},y)P  ~,~\, \\
P^{\pr}:~~\Sigma_1(x_{\mu},-y^{\pr})&=&-P^{\pr} \Sigma_1(x_{\mu},y^\pr)P^{\pr}  ~,~\,\\
P^{\pr}:~~\Sigma_2(x_{\mu},-y^{\pr})&=&P^{\pr} \Sigma_2(x_{\mu},y^\pr)P^{\pr}  ~,~\,\\
P^{\pr}:~~\Sigma_3(x_{\mu},-y^{\pr})&=&P^{\pr}
\Sigma_3(x_{\mu},y^\pr)P^{\pr}~,~\, \eeqa where $\Sigma_1$ gives the
bi-doublet Higgs fields appearing in the 4-D minimal left-right
model while $\Sigma_2$ and $\Sigma_3$ give the $SU(2)_L$ and
$SU(2)_R$ triplet Higgs fields. We can introduce in the bulk the
mixing between different generations
 which is different from the case of gauge-Higgs unification scenario
(because Higgs fields from vector supermultiplets can not mix between
different generations)
\beqa
{\cal L}^{(5)}&=&\sum\limits_{i,j}y_{1ij}^{(5)}X_{ia}^TC \({\Sigma}_1\)^a_bY_j^b
   +\sum\limits_{i,j}y_{2ij}^{(5)}L_{ia}^TC\({\Sigma}_1\)^a_bN_j^b \nonumber \\
&& +\sum\limits_{i,j}y_{3ij}^{(5)}L_{ia}^TC\({\Sigma}_2\)^{ab}L_{jb}
   +\sum\limits_{i,j}y_{4ij}^{(5)}\(N^T\)^a_iC\({\Sigma}_3\)_{ab}N_{jb} \, ,
\eeqa
where $i,j$ are family indices and $a,b$ are group indices.
The decomposition of $SU(4)_W$ adjoint representation in terms of
$SU(2)_L$,$SU(2)_R$ and $U(1)_Z$ with parity assignments of
$\Sigma_i$ is given by
\small
\begin{eqnarray}
(P,P^{\pr}):\bf{15}(\Sigma_1) &=& (\bf{3,1})_0^{+,-}\oplus
  (\bf{1,3})_0^{+,-}\oplus (\bf{2,\bar{2}})_2^{+,+} \oplus
  (\bf{\bar{2},2})_{-2}^{+,+}\oplus (\bf{1,1})_0^{+,-} ~,~\, \nonumber \\
(P,P^{\pr}):\bf{\bar{10}}(\Sigma_2)&=& (\bf{3,1})_0^{+,+}\oplus
  (\bf{1,3})_0^{+,+}\oplus (\bf{2,\bar{2}})_0^{+,-} \nonumber ~,~\,\\
(P,P^{\pr}):\bf{10}(\Sigma_3)&=& (\bf{3,1})_0^{+,+}\oplus
   (\bf{1,3})_0^{+,+}\oplus (\bf{2,\bar{2}})_0^{+,-} ~.~\,
\end{eqnarray}
\normalsize
The four-dimensional effective theory can be obtained by
integrating out the heavy modes which give explicitly the Yukawa coupling
(the subscripts of the Higgs fields denote the representation in
$SU(2)_L \times SU(2)_R$)
\begin{eqnarray}
{\cal L}^{(4)}&=&\sum\limits_{i,j}y_{1ij}^{(4)}Q_i^TC{\Sigma}_{2£¬2}^{(4)}Q_j^c
  +\sum\limits_{i,j}y_{2ij}^{(4)}L_i^TC{\Sigma}_{2£¬2}^{(4)}L_j^c
  \nonumber \\
&& +\sum\limits_{i,j}y_{3ij}^{(4)}L_i^TC{\Sigma}^{(4)}_{3£¬1}L_j
+\sum\limits_{i,j}y_{4ij}^{(4)}(L_i^c)^TC{\Sigma}^{(4)}_{1£¬3}L_j^c+h.c.~.~\,
\end{eqnarray}
The neutrino masses can be generated through type-II see-saw mechanism
via the Higgs triplets.  The extra triplets from the symmetric Higgs fields
couple with the much heavier mirror fermions which can be integrate out
in low energy effective theory.

\item[(ii)] The second possibility is to locate matter on the
$O^{\pr}$-brane at $y=\pi R/2$. Since the gauge symmetry preserves
only for $SU(2)_L \times SU(2)_R \times U(1)_Z$, we need to
introduce $O^\pr$-brane Higgs fields including bi-doublets,
triplets and gauge singlet to break the residue gauge symmetry to
the SM gauge group at the $O^\pr$-brane. Because the $U(1)_Z$
anomalies can not be cancelled with the matter content of the
minimal left-right model, we must introduce new mirror fermions to
cancel the anomaly. Besides, we must introduce gauge singlet Higgs
field to break the $U(1)_Z$. The Yukawa coupling on the
$O^{\pr}$-brane is similar
 to the minimal left-right model except that we must include
the mirror fermions. Further, this scenario can realize gauge
coupling unification but not the unification of matter content
and Higgs.

\item[(iiI)] The third possibility is to put matter fields on the $O$-brane
at $y=0$. Because the gauge symmetry is preserved on the $O$ brane,
we must fit the matter content into $SU(4)_W \times U(1)_{B-L}$
representations in order to give an explanation for unification. We
can introduce bulk Higgs fields or brane Higgs fields. For bulk
Higgs fields, we can introduce $SU(4)_W$ invariant Yukawa
interactions localized on the $SU(4)_W$ invariant $O$-brane. Then we
must specify the transformation properties of the quark and lepton
fields under $Z_2 \times Z_2^\pr$. The parity $P$ under $Z_2$ must
be plus while the parity $P^\pr$ under $Z_2^\pr$ can be determined
by requiring the operator on $(0,\pi R)$ branes to transform
covariantly under $Z_2^\pr$. That is, due to the identification of
$(0,\pi R)$ brane under $Z_2^\pr$, we must specify the
transformation of the matter fields to ensure that the operator on
the two branes are correlated by $Z_2^\pr$. The assignment of the
$P^\pr$ quantum number has four possibilities
\begin{eqnarray}
 && P^\pr(Q ,\bar{Q^c},\bar{Q},Q^c)=\pm(+,-,-,+) \, , \label{p1} \\
 && P^\pr(Q ,\bar{Q^c},\bar{Q},Q^c)=\pm(+,-,+,-) \, . \label{p2}
\end{eqnarray}
We can introduce as in the bulk fermion case the adjoint Higgs
fields $\Sigma_1(x_\mu,y)$ and the symmetric Higgs fields
$\Sigma_2(x_{\mu},y)$ and $\Sigma_3(x_\mu,y)$ with opposite $P^\pr$
parity assignment. Corresponding to the parity assignment in Eqs.
(\ref{p1}) and  (\ref{p2}), we obtain respectively \small
\begin{eqnarray*}
&& P^\pr(X_{\bf 4}Y_{\bf \bar{4}}\Sigma_1)=+,P^\pr(L_{\bf 4}N_{\bf
   \bar{4}}\Sigma_1)=+,P^\pr(L_{\bf 4}L_{\bf 4}\Sigma_2)=+,P^\pr(N_{\bf
   \bar{4}}N_{\bf \bar{4}}\Sigma_3)=+ ~,~\, \\
&& P^\pr(X_{\bf 4}Y_{\bf \bar{4}}\Sigma_1)=-,P^\pr(L_{\bf 4}N_{\bf
  \bar{4}}\Sigma_1)=-,P^\pr(L_{\bf 4}L_{\bf 4}\Sigma_2)=+,P^\pr(L_{\bf
  \bar{4}}L_{\bf \bar{4}}\Sigma_2)=+ ~.~\,
\end{eqnarray*}
\normalsize
So the Yukawa coupling can be written as
\beqa
{\cal L}_5&=&( \delta(y)\pm\delta(y-\pi R))
   \sum\limits_{i,j}(y^f_{1ij}X_i^TC\Sigma_1Y_j+y^f_{2ij}L_i^TC\Sigma_1N_j) \nonumber \\
&& +( \delta(y)+\delta(y-\pi R))\sum\limits_{i,j}(y^f_{3ij}L_i^TC\Sigma_1L_j
   +y^f_{4ij}N_i^TC\Sigma_1N_j) ~,~\,
\eeqa
where $\pm$ correspond to  Eqs. (\ref{p1}) and (\ref{p2}), respectively.
We get the low energy effective theory by integrating the $y$
coordinate. The zero modes of the Yukawa coupling is same as the
minimal left-right model obtained from the bulk fermion cases.
This scenario also has the $U(1)_Z$ anomaly in low energy effective
theory.

It is also possible for the $SU(4)_W$ representation Higgs to lie on
the $O$-brane. Such a possibility is almost identical to the case of
 the ordinary 4-D unification scenario which will be discussed later
 except that the adjoint Higgs which is used to break the gauge symmetry
 from $SU(4)_W$ to $SU(2)_L{\tm}SU(2)_R{\tm}U(1)_Z$ is eliminated.
\end{itemize}

So we can see that in inner automorphism orbifolding symmetry
breaking cases, the most economical way to have left-right
unification is to introduce bulk fermions and bulk Higgs fields.
However, such cases have the $U(1)_Z$ anomalies in the zero modes.

     As mentioned previously,the gauge anomaly will not cancel after orbifold projection.
The relevant discussions on gauge anomaly cancelation in orbifold
was discussed in ref \cite{nima,sssz,pr,lee} etc. In \cite{nima},
the $U(1)$ gauge anomaly in five dimensional theories compactified
on $S^1/Z_2$ with one unit charge bulk fermion was showed to be
lived in the orbifold fix point. The anomaly has the form: \beqa
\pa_M J^M&=&\f{1}{2}\(\delta(y)+\delta(\pi R-y)\){\cal Q}(x_{\mu},y)
\eeqa
 with
 \beqa
{\cal
Q}(x_\mu,y)=\f{g_5^2}{16\pi^2}F_{\mu\nu}(x_\mu,y)\tilde{F}^{\mu\nu}(x_\mu,y)
 \eeqa
 is proportional to the four dimensional chiral anomaly from a charged Dirac
 fermion in the external gauge potential $A_\mu(x_\mu,y)$.
 The current $J^M$ is the five dimensional fermion current
 \beqa
J^M=\bar{\Psi} \Gamma^M \Psi
 \eeqa

 In ref.\cite{sssz},the gauge anomaly was shown to be present in orbifold $S^1/(Z_2\times
 Z_2)$ even in the absence of an anomalous spectrum of zero
 modes.However it was found in \cite{pr} that the theory with a
 single 5D Dirac fermion without anomalous zero modes is by itself non-anomalous.

    In our case, according to the assignments of the fermion parity
 under the orbifold projection(that is,there are fermionic zero modes after projection),
 the anomalous structure for $U(1)_{B-L}$ resemble that of the case in
 ref.\cite{nima}.It can easily be seen that the anomaly in 4-D
 effective theory cancels,so we need not worry about the anomaly for
 $U(1)_{B-L}$.

  The anomaly structure for orbifold broken gauge groups was first discussed in
  \cite{lee}.The bulk fermions(contain both fundamental and anti-fundamental
   representation for $SU(4)_W$ with flipped parity assignments with respect to $P^{\prime}$ )
   give rise to the localized
  gauge anomalies for all gauge components of the five dimensional
  vector current\footnote{In fact in our case,${\cal Q}^a(A)$ and ${\cal Q}^i(A)$ vanish.}:
\beqa (D_MJ^M)^a(x_\mu,y)&=&\delta(y)\[{\cal Q}^a(A)+{\cal Q}^a(X)\]
\\     (D_MJ^M)^i(x_\mu,y)&=&\delta(y)\[{\cal Q}^i(A)+{\cal Q}^i(X)
\]\nonumber\\
   (D_MJ^M)^B(x_\mu,y)&=&\delta(y)\[{\cal Q}^B_{+}(A)+{\cal Q}^B_{-}(A)\]
+\delta(y){\cal Q}^B(X) \nonumber\\
(D_MJ^M)^{\hat{a}}(x_\mu,y)&=&\delta(y){\cal Q}^{\hat{a}}(X)
\nonumber
 \eeqa
here the superscript $a,i$ represent the two unbroken non-abelian
gauge groups $SU(2)_L$ and $SU(2)_R$ generators; the superscript $B$
represent the unbroken abelian gauge group $U(1)_Z$ from the
diagonal $SU(4)_W$; the superscript $\hat{a}$ denote the broken
generators for the previous $SU(4)_W$.${\cal Q}^M$ is again the four
dimensional gauge anomaly relate to the anomaly via: \beqa {\cal
Q}^M &\propto&
\f{1}{16\pi^2}\sum\limits_{N,L}Tr(\{T^M,T^N\}T^L)F_{\mu\nu}^N\tilde{F}^{L\mu\nu}\nonumber\\
&=&\sum\limits_{N,L}\f{1}{32\pi^2}D^{MNL}F_{\mu\nu}^N\tilde{F}^{L\mu\nu}\eeqa
here $N,L$ run through all the $SU(4)_W$ generation indices and
$D^{MNL}$ denotes the symmetrized trace.

To cancel the local anomaly,we need to place localized fermions in
the $y=\f{\pi R}{2}$ brane so as that the total anomaly can be
canceled by Chern-Simons term.That is, we introduce localized
fermions in $SU(3)_c\times SU(2)_L\times SU(2)_R \times U(1)_Z$
representation:
$${\bf (3,\bar{2},1)_{-1},(3,1,\bar{2})_{+1},(1,\bar{2},1)_{-1},(1,1,\bar{2})_{+1}}$$ on the
$y=\f{\pi R}{2}$ brane.

So the bulk anomaly changed into the form: \beqa
(D_MJ^M)^a(x_\mu,y)&=&\delta(y){\cal
Q}^a(X)+\[\delta(y)-\delta(y-\f{\pi R}{2})\]{\cal Q}^a(A)
\\     (D_MJ^M)^i(x_\mu,y)&=&\delta(y){\cal Q}^i(X)
+\[\delta(y)-\delta(y-\f{\pi R}{2})\]{\cal Q}^i(A)\nonumber\\
   (D_MJ^M)^B(x_\mu,y)&=&\[\delta(y)-\delta(y-\f{\pi R}{2})\]\[{\cal Q}^B_{+}(A)+{\cal Q}^B_{-}(A)\]
+\delta(y){\cal Q}^B(X) \nonumber\\
(D_MJ^M)^{\hat{a}}(x_\mu,y)&=&\delta(y){\cal Q}^{\hat{a}}(X)
\nonumber
 \eeqa

 We can introduce the Chern-Simons term to cancel the gauge anomaly.
We introduce the deformed Chern-Simons 5-form $Q_5[A^MT^M]$ in the
action: \beqa {\cal L}_{CS}=-\f{1}{48\pi^2}u(y)
Tr\(AdAdA+\f{3}{2}A^3dA+\f{3}{5}A^5\) \eeqa with $u(y)$ a parity odd
function.
  Under gauge transformation $\delta A=d \omega +[A,\omega]$, from
  the variation of the Lagrangian we get the five dimensional
  covariant gauge current:
\beqa \(D_M
J^M\)^{O=(a,i,B)}&=&-\f{1}{32\pi^2}\[\delta(y)-\delta(y-\f{\pi
R}{2})\]
\sum\limits_{P,Q=(a,i,B)}D^{OPQ}F^{P}_{\mu\nu}F^{Q\mu\nu}\nonumber\\
&-&\f{1}{32\pi^2}\delta(y)\sum\limits_{P,Q=(\hat{a})}D^{OPQ}F^{P}_{\mu\nu}F^{Q\mu\nu}\\
\(D_M
J^M\)^{O=\hat{a}}&=&-\f{1}{32\pi^2}\delta(y)\sum\limits_{P=\hat{a},Q=(a,i,B)}D^{OPQ}F^{P}_{\mu\nu}F^{Q\mu\nu}
\eeqa
 So we can see that the Cherm-Simons contributions cancel exactly the gauge
 anomalies.

 In general, the $U(1)_Z$ is anomalous in four dimension. We must
 introduce the localized brane fermions and Chern-Simons term to
 eliminate the gauge anomaly. So we want to seek new ways in
 orbifolding breaking to eliminate the anomalous $U(1)_Z$.
\subsection{Five-dimensional outer automorphism orbifold breaking}
  It is well known that inner automorphism orbifolding breaking with $Z_n$ action can not reduce
  the rank of the gauge groups.So we seek to use outer automorphism orbifolding breaking
  mechanism \cite{hebecker,quiros} to
  eliminate the rank of the group. Outer automorphisms are structure
  constant preserving linear transformations of generators which
  cannot be written as group conjugations.As an example, complex conjugation which
  preserve the structure constant can not be written as a
  conjugation by group elements.Such type of orbifolding procedure in general
  reduce the rank of the group.

 We proposed that the left-right unification  model can be broken to
  the minimal left-right model by the rank-reducing outer
  automorphism  orbifolding breaking mechanism
  through the breaking pattern $SU(4)\to SO(4)$. We know that
$SO(4)\sim SU(2) \times SU(2)$ where we can identify the two
$SU(2)$ as $SU(2)_L$ and $SU(2)_R$. In this way, we
eliminate the anomalous $U(1)_Z$ appeared in the inner automorphism
orbifolding breaking. So the symmetry breaking pattern reads
 \beqa
 SU(4)_{w}{\tm}U(1)_{B-L} &\ra& SO(4){\tm}U(1)_{B-L}
 \sim SU(2)_L{\tm}SU(2)_R{\tm}U(1)_{B-L} ~.~\,
 \eeqa
In this case we also consider the five-dimensional spacetime ${\cal
M}_4{\times}S^1/(Z_2{\times}Z_2)$. As in the inner automorphism
case, there are two inequivalent 3-branes $O$ and $O^{\pr}$
corresponding to $y=0$ and $y=\pi R/2$, respectively. The $SU(4)_W$
gauge theory is defined on the orbifold $S^1/(Z_2{\times}Z_2)$ with
\beqa
T^a A^a_{\mu}(x_{\mu},-y)&\sim& -(T^a)^*A_{\mu}^a(x_{\mu},y) ~,~\, \\
T^a A_5^a(x_{\mu},-y)&\sim& (T^a)^* A_{5}^a(x_{\mu},y) ~,~\, \\
T^a A^a_{\mu}(x_{\mu},-y^{\pr})&\sim& P^{\pr}(T^a)P^{\pr}
A_{\mu}^a(x_{\mu},y^{\pr}) ~,~\,\\
T^a A_5^a(x_{\mu},-y^{\pr})&\sim& -P^{\pr}(T^a)P^{\pr}
A_{5}^a(x_{\mu},y^{\pr}) ~.~\, \eeqa We give positive (negative)
parity to gauge fields corresponding to the imaginary antisymmetric
(real symmetric) generators of $SU(4)$. With such an assignment the
$SO(4)$ gauge fields have positive parity while other gauge fields
have negative parity. We can choose the parity assignment in terms
of the fundamental representation of $SU(4)_W$ for $P^{\pr}$: \beqa
P^{\pr}=diag(+1,+1,+1,+1) ~.~\,\eeqa So we know that the gauge
symmetry is preserved on $O^{\pr}$ brane while is broken on the $O$
brane to $SO(4){\sim}SU(2)_L{\times}SU(2)_R$. The $Z_2$ assignment
for the matter content in the bulk is defined as \beqa
\psi(x_{\mu},-y)=\lambda_R\psi(x_{\mu},y) ~,~\,\eeqa with
$\la_R=\la^{\da}_R=\la_R^{-1}$ being a matrix acting on the
representation indices of $\psi$. As for the fermions in the
representation $R$ of the group $G$, the requirement for the
fermion-gauge boson coupling to be invariant under orbifold action
implies that $\la_R$ should satisfy the condition \beqa
\la_RT_R^A\la_R=\Lambda^{A}_{B}T_R^B \, . \eeqa For such a complex
conjugate outer automorphism breaking, the former requirements have
the form \beqa \la_RT_R^A\la_R=-(T_R^A)^* \, . \eeqa Such an
identity requires the representation to be real. As we can always
choose $R=r\oplus\bar{r}$ where $r$ is a non-real representation
with generators \beqa
T_R^A=\left(\bea{cc}T^A_r&0\\0&-(T_r^A)^*\eea\right) ~,~\,\eeqa we
can choose the form of $\la_R$ that satisfies the former
requirements \beqa \la_{R}=\(\bea{cc} 0&{\bf 1}_r\\ {\bf 1}_r&0
\eea\) \, . \eeqa We know that the representation of the matter
content require a $r{\leftrightarrow} \bar{r}$ symmetry to guarantee
the action to be invariant under orbifolding. So we see that the
original bulk theory must be vector-like (with respect to the group
we wish to act on by outer automorphism). We do not specify the way
for the matter to fill the ${\bf 4}$ and ${\bf \bar{4}}$ now and
just assume them to be ${\bf 4} \sim (M_i)$, ${\bf 4} \sim (R_i)$
and ${\bf \bar{4}} \sim (N_i)$ ${\bf \bar{4}} \sim (T_i)$
($i=0,1,2,3$). The matter content in original ${\bf 4} \oplus{\bf
\bar{4}}$ of $SU(4)_W$ transforms in ${\bf 4} \oplus{\bf 4}$ in
$SO(4)$. The parity of the states are \beqa
\la_{4\oplus4}=\(\bea{cc} 0&{\bf 1}_4\\ {\bf 1}_4&0 \eea\)
~.~\,\eeqa After diagonalization, we obtain the eigenvalues $\pm1$
with 4 even states and 4 odd states, respectively. The corresponding
eigenvectors are $(e_i, \pm e_i)$ with $e_i$ being a unit vector in
$i$th direction. We get the parity assignment in terms of the
combination (similar results for lepton sector) \beqa \left(
\bea{cc} {\bf ~1}_4&{\bf ~1}_4\\ {\bf ~1}_4&{\bf -1}_4 \eea
\right)\left(\bea{c} M_i\\N_i\eea\right) (x_{\mu},-y){\sim} \left(
\bea{cc} {\bf 1}_4&\\ &{\bf  -1}_4
 \eea \right)\left( \bea{cc} {\bf ~1}_4&{\bf ~1}_4\\
 {\bf ~1}_4&{\bf -1}_4
\eea \right)\left(\bea{c} M_i\\N_i\eea\right)(x_{\mu},y)~.~\, \eeqa
Then we can see that the combination $M_i-N_i$ has negative parity
and is projected out. The zero mode of the combination $M_i+N_i$ is
a $SO(4)$ vector which survives the projection. We use the method
similar to the Lorentz group to determine the two $SU(2)$
transformations of the fermions. We denote $M_i+N_i$ as a $SO(4)$
vector and multiply the sigma matrix $\sigma_i$ to get the two
$SU(2)$ indices $\alpha,\dot{\beta}$: \beqa
\left(M+N\right)_i\sigma^i_{\al\dot{\beta}}=\left( \bea{cc}
 M_0+M_3+N_0+N_3&M_1-iM_2+N_1-iN_2\\
 M_1+iM_2+N_1+iN_2&M_0+M_3-N_0-N_3\eea\right)~.~\,
\eeqa
These two  indices transform as two SU(2), respectively.
We denote them as
\beqa
\left(M+N\right)_i\sigma^i_{\al\dot{\beta}}{\sim}\left(\bea{c}u_L\\
 d_L\eea\right)_{\al}{\otimes}(\bea{c}u_L^c~~d_L^c\eea)_{\dot{\bt}}~.~\,
\eeqa
In this way we identify the zero-mode fermions as the matter
content in the minimal left-right model. This identification
of the matter content is different from the case in the inner
automorphism breaking mechanism.

The Yukawa couplings are introduced in the bulk with the bulk Higgs
fields in the adjoint and symmetric 10-dimensional representation
of the $SU(4)_W$. The parity for the adjoint Higgs has a relative
minus sign with respect to the gauge fields
\beqa
(T^a) \Sigma^a(x_{\mu},-y)&\sim& -(T^a)^*\Sigma^a(x_{\mu},y)~.~\,
\eeqa
The decomposition of the adjoint Higgs with respect to the two $SU(2)$
is given by
\small
\beqa
(P,P^{\pr}):\bf{15}(\Sigma_1) &=& (\bf{3,1})^{-,+}\oplus
 (\bf{1,3})^{-,+}\oplus (\bf{2,\bar{2}})^{+,+} \oplus
 (\bf{\bar{2},2})^{+,+}\oplus (\bf{1,1})^{+,+} ~.~\,
\eeqa \normalsize In this case, an extra singlet Higgs field is
preserved after projection. It can be used to break the left-right
parity \cite{dparity1,dparity2} in the remaining left-right model.

Similar to the parity assignment of the bulk fermions, the
parity of the symmetric Higgs fields are also non-diagonal.
We must include both ${\bf 10}$ and ${\bf \bar{10}}$ Higgs
fields denoted  respectively as $\Sigma_2(x_\mu,y)$ and $\Sigma_3(x_\mu,y)$
\beqa
\( \bea{c} \Sigma_2\\ \Sigma_3 \eea\)(x_\mu,-y)=\(\bea{cc}0&{\bf 1}_{10}
 \\{\bf 1}_{10}&0\eea \)\( \bea{c} \Sigma_2\\ \Sigma_3 \eea\)(x_\mu,y)~.~\,
\eeqa
The non-diagonal matrix has eigenvalues $\pm1$ with 10 positive and 10
negative ones. After projection, only one $SO(4)$ 10-dimensional Higgs
field $\Delta$ is kept. The symmetric 10-dimensional representation for
$SO(4)$ can be decomposed in term of $SU(2)_L \times SU(2)_R$ as
 \beqa
(P,P^\pr):\Delta(\bf{10})=\bf{(3,1)}\oplus\bf{(1,3)}\oplus\bf{(2,2)}~.~\,
 \eeqa
 The invariant Yukawa couplings in five dimensions must be symmetric under
 $r{\leftrightarrow}\bar{r}$ and is given by
\small
\beqa {\cal L}_5=\sum\limits_{i,j=1}^3(y_{1ij}M_i^TC\Sigma_1N_j
 +y_{2ij}R_i^TC\Sigma_1T_j+ y_{3ij}T_i^TC\Sigma_2T_j+y_{3ij}R_i^TC\Sigma_3R_j)~.~\,
\eeqa
\normalsize
The invariance under $P$ transformation requires the Yukawa coupling
of the 10-dimensional Higgs to be identical. The low energy effective
theory (with zero mode of Higgs and fermion KK modes) of the Yukawa
coupling part is similar to the inner automorphism breaking scenario
except that there is an additional contribution of bi-doublet coupling
from the symmetric Higgs fields given by
 \begin{eqnarray}
{\cal L}^{(4)}&=&\sum\limits_{i,j}y_{1ij}^{(4)}Q_i^TC{\Sigma}_{2£¬2}^{(4)}Q_j^c
  +\sum\limits_{i,j}y_{2ij}^{(4)}L_i^TC{\Sigma}_{2£¬2}^{(4)}L_j^c+\sum\limits_{ij}y_{3ij}Q_i^TC S
  Q_j^c+\sum\limits_{ij}y_{4ij}L_i^TC S
  L_j^c
 \nonumber \\&& +\sum\limits_{i,j}y_{5ij}^{(4)}L_i^TC{\Sigma}^{(4)}_{3£¬1}L_j
+\sum\limits_{i,j}y_{5ij}^{(4)}(L_i^c)^TC{\Sigma}^{(4)}_{1£¬3}L_j^c
  +\sum\limits_{i,j}y_{5ij}^{(4)}L_i^TC{\Sigma}_{2£¬2}^{(4)}L_j^c ~.~\,
 \end{eqnarray}
When we introduce the fermions on the $O^\prime$ brane and the Higgs
fields in the bulk, we need to specify the transformation law for
the fermions to get $SU(4)_W$ invariant interactions. The
transformation of fermions under parity $P$ is non-trivial and we
cannot assign diagonal parity matrix for the fermions as in the
inner automorphism case. The parity assignment for the fermions is
similar as for the bulk fermions. The parity for the fermions (same
for the lepton sector) is given by
  \beqa
 \left(\bea{c} M_i\\N_i\eea\right) (x_{\mu},-\f{\pi R}{2})=\pm \left( \bea{cc}0& {\bf 1}_4 \\
 {\bf  1}_4&0\eea \right)\left(\bea{c} M_i\\N_i\eea\right)(x_{\mu},\f{\pi R}{2} ) ~.~\,
 \eeqa
   So the Yukawa coupling with the adjoint and symmetric Higgs fields
reads \small \beqa {\cal L}_5&=&( \delta(y-\pi R/2)+\delta(y+\pi
R/2))\sum\limits_{i,j}(y^f_{1ij}M_i^TC\Sigma_1N_j+y^f_{2ij}R_i^TC\Sigma_2T_j)
\nonumber \\
&&+( \delta(y-\pi R/2)+\delta(y+\pi R/2))
\sum\limits_{i,j}(y^f_{3ij}R_i^TC\Sigma_2R_j+y^f_{3ij}T_i^TC\Sigma_3T_j)
~.~\, \eeqa \normalsize
   Here the parity of the bulk Higgs is
identical to the previous bulk fermions.The Yukawa couplings to give
the Majorana masses for the neutrinos are the same for left and
right parts because of the previous reflection type transformation
requirement. The low energy effective theory with zero mode of the
Higgs fields is same as in the bulk cases.

We can also locate the fermions on the brane $O$. The gauge symmetry
is preserved to be $SU(2)_L \times SU(2)_R$ on this brane. In this
case, the spectrum can be similar to the minimal left-right model
but with the virtue of gauge symmetry unification. It avoids the
problem of heavy-light Higgs splitting but does not have fermion and
Higgs unification.

 The anomaly cancelation in outer automorphism orbifold broken case is
 different to the case of inner automorphism broken. The $U(1)_Z$
 appears in previous section is no longer present
 in the four dimensional low energy effective theory.

 The advantage of outer automorphism orbifold broken is most obvious in
 $S^1/Z_2$ orbifolding case(or similar case for $P=P^{\prime}=(+,+,-,-)$
 in $S^1/(Z_2\times Z_2)$ orbifolding ).
 In this case,no localized anomalies related
 to broken generators of $SU(4)_W$ appear in the 5D vector
 current.At the same time, because of the eliminated $U(1)_Z$,the
 four dimensional anomalies related to $SU(2)_L$ and $SU(2)_R$
 vanish.Thus the theory is non-anomalous.
We can see that the employment of the rank-reducing outer
automorphism orbifolding symmetry breaking mechanism can eliminate
the anomalous $U(1)_Z$, and thus the theory is free from the
Chern-Simons terms and new localized fermions.

 In most general case ($P\neq P^{\prime}$) of $S^1/(Z_2\times Z_2)$ orbifolding with
 outer automorphism broken mechanism, the localized anomaly in
 general can not be eliminated automatically.
The form of the anomaly in our case can be written: \beqa
(D_MJ^M)^a(x_\mu,y)&=&\delta(y)\[{\cal Q}^a(X)+{\cal Q}^a(A)\]
\\     (D_MJ^M)^i(x_\mu,y)&=&\delta(y)\[{\cal Q}^i(X)
+{\cal Q}^i(A)\]\nonumber\\
   (D_MJ^M)^{\hat{B}}(x_\mu,y)&=&\delta(y)\[{\cal Q}^{\hat{B}}_{+}(A)+{\cal Q}^{\hat{B}}_{-}(A)\]
+\delta(y){\cal Q}^{\hat{B}}(X) \nonumber\\
(D_MJ^M)^{\hat{a}}(x_\mu,y)&=&\delta(y){\cal Q}^{\hat{a}}(X)
\nonumber
 \eeqa
here $\hat{B}$ denote the broken $U(1)_Z$ generator.The cancelation
of the anomaly is identical to that of the inner automorphism
orbifolding mechanism.

 We can see that due to the elimination of the extra $U(1)_Z$, it is possible for the outer automorphism broken
 case to be non-anomalous without the introduction of Chern-Simons term and localized
 fermions.

\subsection{Symmetry breaking in four dimensions via Higgs mechanism}
In this section we discuss our model in four dimensions with the Higgs mechanism
for symmetry breaking. This framework provides the most direct extension of
the left-right model and does not have the arbitrariness which appears in
the orbifold projection in five dimensions.

The matter content is shown in Sec. 2.1.  The gauge symmetry breaking
is through the Higgs mechanism. We introduce 15-dimensional adjoint
representation Higgs fields $\Sigma,\Phi$ with vanishing $U(1)_{B-L}$
charge, 10-dimensional symmetric Higgs field $\Delta$ with $U(1)_{B-L}$
charge $2$ and gauge singlet $S$. The Higgs potential is given by
\begin{eqnarray}
V(\Sigma_i^j,\Phi_{m}^n,\Delta_{k,l},S)=V_{\Sigma}+V_{\Phi}+V_{\Delta}+V_{S}+V_{cross}~,~\,
\end{eqnarray}
where
\small
\begin{eqnarray}
V(\Sigma)&=&-m_1^2Tr(\Sigma^2)+\lambda_1[Tr(\Sigma^2)]^2+\lambda_2Tr(\Sigma^4)  ~,~\,\\
V(\Phi)&=&-m_2^2Tr(\Phi^2)+\lambda_3[Tr(\Phi^{2})]^2+{\lambda_4}Tr(\Phi^{4}) ~,~\,\\
V(\Delta)&=&-m_3^2Tr(\Delta^{\dagger}\Delta)+{\lambda_5}[Tr(\Delta^{\dagger}\Delta)]^2
            +{\lambda_6}Tr(\Delta^{\dagger}\Delta\Delta^{\dagger}\Delta) ~,~\,\\
V(S)&=&-m_4^2(S^{\dagger}S)+\lambda_7(S^{\dagger}S)^2 ~,~\, \\
V_{cross}&=&\chi_1(\Delta^{\dagger}\Delta)Tr(\Sigma^2)
 +\chi_2(\Delta^{\dagger}\Delta)Tr(\Phi^2) +\chi_3(S^{\dagger}S)Tr(\Sigma\Sigma) \nonumber \\
&& +\chi_4(S^{\dagger}S)Tr(\Phi\Phi)+\chi_5(S^{\dagger}S)(\Delta^{\dagger}\Delta)
   +\chi_6Tr(\Sigma^2\Phi^2)+\chi_7(\Phi\Sigma\Phi\Sigma) ~.~\,
\end{eqnarray}
\normalsize
Here we impose a discrete symmetry as ${\Phi}{\leftrightarrow}-{\Phi}$
to eliminate various cubic terms of Higgs fields. The charge assignment
for the adjoint Higgs $\Phi,\Sigma$ and the symmetric tensor Higgs field
$\Delta$ reads
\begin{eqnarray}
Q(\Phi)=\left( \begin{array}{cccc}
 0& 1& 0&1\\ -1&0& -1&0 \\ 0&1 &0&1\\ -1& 0&-1 &0 \\
 \end{array} \right)~~,~~
 Q(\Delta)=\left( \begin{array}{cccc}
  0& -1& 0&-1\\ -1&-2& -1&-2 \\ 0&-1&0&-1 \\ -1& -2&-1 &-2 \\
\end{array} \right)~.~\,
\end{eqnarray}
The first step of symmetry breaking from
$SU(4)_W \times U(1)_{B-L}\rightarrow SU(2)_L\times SU(2)_R \times
U(1)_{B-L} \times U(1)_Z$ is accomplished by the Higgs field $\Phi$.
We can find values of non-zero $<\Phi>$ from minimizing the Higgs
potential \cite{LLF}. The minimum of the potential $V_{\Phi}$ can
be written as
\begin{eqnarray}
<\Phi>=v\left( \begin{array}{cccc}
 1& & &\\ &1& & \\ & &-1& \\ & & &-1 \\ \end{array} \right)\, ,
\end{eqnarray}
with
\begin{eqnarray}
v^2=\frac{m_1^2}{8\lambda_1+\lambda_2} \, .
\end{eqnarray}
We can parameterize $\Phi$ as
\begin{eqnarray}
&& \Phi-<\Phi>= \left( \begin{array}{cc}
  [H_3]^i_j&H^i_j\\ (H^{\dagger})^i_j&[H_3]^i_j\\ \end{array}\right) \, ,\\
&& <\Phi>_i^j=y_i\delta_i^j  \, .
\end{eqnarray}
The kinetic term for the adjoint Higgs fields $\Phi$ reads
\begin{eqnarray}
D_{\mu}\Phi&=&\partial_{\mu}\Phi+ig_4(A_{\mu}\Phi-\Phi A_{\mu})
 =\partial_{\mu}\Phi+ig_4(T^a\Phi-\Phi T^a)A^{a}_{\mu} \, .
\end{eqnarray}
After symmetry breaking, the mass term of gauge bosons reads
\begin{eqnarray}
&& [A_{\mu},<\Phi>] =(A_{\mu})_i^j<\Phi>_j^k-<\Phi>_i^j(A_{\mu})_j^k
   =(A_{\mu})_i^k(y_k-y_i) \, , \\
&& g_4^2Tr\{[A_{\mu},<\Phi>]^{\dagger}[A^{\mu},<\Phi>]\}
  =g_4^2\sum\limits_{i,j} \{(A_{\mu})_j^i\}^*(A^{\mu})_j^i(y_j-y_i)^2 \, .
\end{eqnarray}
The gauge fields components are identified as
\small
\begin{eqnarray}
&& (A_{\mu})=\frac{1}{2}\left(\begin{array}{ccc}
   (A_{\mu})^L+\frac{\sqrt{2}}{2}B_{\mu}^{Z}&\vline&\sqrt{2}A_{\mu}^H\\
  &\vline&\\ \hline &\vline&\\
  \sqrt{2}(A_{\mu}^H)^{\dagger}&\vline&(A_{\mu})^R-\frac{\sqrt{2}}{2}B_{\mu}^{Z}
  \end{array} \right) \, , \\
&& (A_{\mu})^L=\left( \begin{array}{cc}
    (A_{\mu}^0)^L&~~\sqrt{2}(A_{\mu}^+)^L\\ \sqrt{2}(A_{\mu}^-)^L&~~-(A_{\mu}^0)^L
    \end{array}\right)  \, , \\
&& (A_{\mu})^R=\left( \begin{array}{cc}
    (A_{\mu}^0)^R&~~\sqrt{2}(A_{\mu}^+)^R\\ \sqrt{2}(A_{\mu}^-)^R&~~-(A_{\mu}^0)^R
     \end{array}\right) \, ,\\
&& (A_{\mu})^H=\left( \begin{array}{cc}
   X_{\mu}^+&~~Y_{\mu}^{++}\\ X_{\mu}^0&~~Y_{\mu}^+ \end{array}\right) \, ,
\end{eqnarray}
\normalsize
where $(A_{\mu})_L$is in the $SU(2)_L$ adjoint representations
$(\bf{3},\bf{1})$, $(A_{\mu})_R$ is in the $SU(2)_R$ adjoint
representation $(\bf{1},\bf{3})$, and $(A_{\mu})^H$ is the bi-doublet
of $SU(2)_L$ fundamental representation and $SU(2)_R$ anti-fundamental
representation $(\bf{2},\bar{\bf{2}})$.

As $\langle\Phi\rangle$ preserves $SU(2)_L{\times}SU(2)_R{\times}U(1)_{Z}$,
the massive gauge bosons have the mass
\begin{eqnarray}
M_X^2=M_Y^2= g_4^2 v^2 \, .
\end{eqnarray}
After the first step of symmetry breaking, we can integrate out
the heavy Higgs modes to induce symmetry breaking in the second step
\begin{eqnarray}
\langle\Delta\rangle=\left( \begin{array}{cccc}
   0& & &\\ &0& & \\ & &v_S& \\ & & &0 \\ \end{array} \right) \, .
\end{eqnarray}
It can be easily seen that the gauge groups $SU(2)_R$, $U(1)_{B-L}$
and $U(1)_Z$ are broken by such vevs.  The kinetic terms for the
adjoint and symmetric Higgs fields $\Sigma,\Delta$ read
\begin{eqnarray}
{\cal L}_{k}=Tr\[(D_{\mu}\Sigma)^{\dagger}(D^{\mu}\Sigma)\]
  +Tr\[(D_{\mu}\Delta^{\dagger})(D^{\mu}\Delta)\] \, ,
\end{eqnarray}
where
\begin{eqnarray}
D_{\mu}\Sigma&=&\partial_{\mu}\Sigma+ig_4[A_{\mu},\Sigma] \, ,\\
D_{\mu}\Delta&=&{\partial}_{\mu}\Delta-ig_4A_{\mu}\Delta-ig_4{\Delta}A_{\mu}
  +i \frac{1}{2}Y_{X} g_{X}A_{\mu}^{X}\Delta \, .
\end{eqnarray}
Now we  get the contributions to the gauge boson masses given by
\small
\begin{eqnarray}
&& g_4A_{\mu}\langle\Delta\rangle +g_4\langle\Delta \rangle A_{\mu}
  -g_XA_{\mu}^{X} \langle\Delta \rangle = \nonumber \\
&& ~~~~~~\f{1}{2}\left(\begin{array}{cccc}
   0&0&0&0\\ 0&0&0 &0\\
   0&0&2g_4[(W_{\mu}^0)^R-\frac{\sqrt{2}}{2}B_{\mu}^{Z}]v_S
      -2g_XA_{\mu}^Xv_S & ~~~~\sqrt{2}g_4(W_{\mu}^+)^Rv_S\\
   0&0& \sqrt{2}g_4(W_{\mu}^-)^Rv_S&0 \end{array} \right)~.~\,
\end{eqnarray}
\normalsize
Here we neglected the terms relevant to $X$ or $Y$ gauge fields
since at low energy we can integrate out them by the equation of
motion and get suppression by order $(v_i/v)^2$ which is very
small. The mass terms for the gauge bosons are given by
\begin{eqnarray}
{\cal L}&=&\frac{1}{4}\left\{4[(W_{\mu}^0)^R
 -\frac{\sqrt{2}}{2}B_{\mu}^{Z}]^2g_4^2v_S^2
 -4g_Xg_4v_S^2(2A_{\mu}^X)[(W_{\mu}^0)^R-\frac{\sqrt{2}}{2}B_{\mu}^{Z}]
  \right. \nonumber \\
&& \left. +4
g_4^2(W_{\mu}^-)^R(W_{\mu}^+)^Rv_S^2+g_X^2v_S^2(2A_{\mu}^X)^2\right\}~.~\,
\end{eqnarray}
\normalsize
We know from the combination
\begin{eqnarray}
{\bf 4 } ~{\otimes} ~{\bf  4^*} ={\bf 15} ~{\oplus} ~{\bf 1}~,~\,
\end{eqnarray}
that the fermions can acquire masses through couplings to $SU(4)_W$
adjoint and singlet Higgs fields. We can introduce $SU(4)_W$ singlet
Higgs field $S$ with vanishing $B-L$ charge to give fermion masses.
We can express the vevs of $S$ as
\begin{eqnarray}
\langle S\rangle =\left( \begin{array}{cccc}
  v_0&&&\\ &v_0&& \\ & &v_0& \\ & & &v_0 \end{array} \right)~.~\,
\end{eqnarray}
Now we can construct the $SU(4)_W$ invariant Yukawa couplings.
The Yukawa couplings including mixing between generations can be
written as \footnote{Note that in principle the Higgs field $\Phi$ can also
couple to the fermions which can give extremely large masses for
fermions and result in almost vector fermions. We can forbid such
couplings by introducing discrete symmetry so as that $\Phi$ and
$\Sigma$ have opposite parity}
\begin{eqnarray}
{\cal L}_{yukawa1}&=&\sum\limits_{ab}y_{1ab}^f(Y_L)^T_{ia}C\Sigma_{j}^i(X_L)_b^j
  +\sum\limits_{ab}y_{2ab}^f(E_L)^T_{ia}C\Sigma_{j}^i(L_L)_b^j\nonumber \\
&&+\sum\limits_{ab}y_3^f(Y_L)^T_{ia}C(X_L)_b^iS
  +\sum\limits_{ab}y_4^f(E_L)^T_{ia}C(L_L)^i_bS+h.c.~.~\,
\end{eqnarray}
The fermions acquire masses after $\Sigma$ develops vevs of the form
(the imaginary part in the vev can be absent to avoid spontaneously
CP broken):
\begin{eqnarray}
\langle\Sigma\rangle=\left( \begin{array}{cccc}
v_5& &v_1-i v_2 & 0\\ 0&v_6& &v_3-iv_4 \\ v_1+i v_2&0 &-v_5& 0\\ 0& v_3+i v_4& 0&-v_6 \\
\end{array} \right)~.~\,
\end{eqnarray}
The vev of $\Sigma$ breaks the symmetry completely into
$U(1)_Q$. The corresponding masses for various gauge bosons can be
obtained from such vevs. From $\langle\Sigma\rangle$ in the above equation and
$A_{\mu}$ in the form of
\small
\begin{eqnarray}
\hspace*{-0.5cm} A_{\mu}=\f{1}{2} \left( \begin{array}{cccc}
(W_{\mu}^0)^L+\frac{\sqrt{2}}{2}B_{\mu}^{Z}& \sqrt{2}(W_{\mu}^+)^L&\sqrt{2} X_{\mu}^+&\sqrt{2}Y_{\mu}^{++}\\
\sqrt{2}(W_{\mu}^-)^L&-(W_{\mu}^0)^L+\frac{\sqrt{2}}{2}B_{\mu}^{Z}& \sqrt{2}\bar{X}_{\mu}^0&\sqrt{2}Y_{\mu}^+ \\
\sqrt{2}X_{\mu}^-& \sqrt{2}\bar{X}_{\mu}^0&(W_{\mu}^0)^R-\frac{\sqrt{2}}{2}B_{\mu}^{Z}&\sqrt{2}(W_{\mu}^+)^R\\
\sqrt{2}Y_{\mu}^{--}&\sqrt{2} X_{\mu}^-&\sqrt{2}(W_{\mu}^-)^R &-(W_{\mu}^0)^R-\frac{\sqrt{2}}{2}B_{\mu}^{Z}\\
\end{array} \right)~,~\,
\end{eqnarray}
\normalsize
we can obatin $A_{\mu}\langle\Sigma\rangle -\langle\Sigma\rangle A_{\mu}$.
Then we get the mass terms for the gauge bosons:
\small
\begin{eqnarray}
\Delta {\cal L}&=&
 \frac{1}{2}g_4^2\left\{\right.
 (v_1^2+v_2^2+v_3^2+v_4^2)\left( [(W_{\mu}^0)^L]^2 +[(W_{\mu}^0)^R]^2 \right)\nonumber \\
&& +2[v_1^2+v_2^2+v_3^2+v_4^2+(v_5-v_6)^2]
   \left[ (W_{\mu}^+)^L(W_{\mu}^-)^L + (W_{\mu}^+)^R(W_{\mu}^-)^R \right] \nonumber \\
&& -2\left( 2(v_1+iv_2)(v_3-iv_4)(W_{\mu}^+)^L(W_{\mu}^-)^R +h.c.\right)\nonumber \\
&& +4(v_1^2+v_2^2+v_3^2+v_4^2)\left(
(\frac{\sqrt{2}}{2}B_{\mu}^{Z})^2
 - \f{1}{2}(W_{\mu}^0)^L (W_{\mu}^0)^R \right)   \nonumber \\
&& +4(v_1^2+v_2^2-v_3^2-v_4^2)[(W_{\mu}^0)^L
   -(W_{\mu}^0)^R][\frac{\sqrt{2}}{2}B_{\mu}^{Z}] \left. \right\} \, .
\end{eqnarray}
\normalsize
So we get the $W_L-W_R$ mixing matrix:
\small
\begin{eqnarray}
\left( \begin{array}{cc}
g_4^2[v_1^2+v_2^2+v_3^2+v_4^2+(v_5-v_6)^2]&-2g_4^2 (v_1-iv_2)(v_3+iv_4)\\
- 2g_4^2 (v_1+iv_2)(v_3-iv_4)& g_4^2 [v_1^2+v_2^2+v_3^2+v_4^2+(v_5-v_6)^2+v_S^2]\\
\end{array} \right)~.~\,
\end{eqnarray}
\normalsize
The mass eigenstates for the gauge bosons are
\begin{eqnarray}
W_1&=&W_L\cos\zeta+W_R\sin\zeta~~,~~W_2=-W_L\sin\zeta+W_R\cos\zeta \, ,
\end{eqnarray}
with masses given by
\begin{eqnarray}
M^2_{W_1}&{\simeq}&g_4^2\frac{4 (v_1^2+v_2^2)(v_3^2+v_4^2)}{
v_S^2}+g_4^2[v_1^2+v_2^2+v_3^2+v_4^2+(v_5-v_6)^2]~,~\,\\
M^2_{W_2}&{\simeq}&g_4^2v_S^2 \, ,
\end{eqnarray}
and the mixing angle $\zeta$ given by
\begin{eqnarray}
\tan{2\zeta}=\frac{4 \sqrt{(v_1^2+v_2^2)(v_3^2+v_4^2)}}{v_S^2} \, .
\end{eqnarray}
The mixings between the neutral components ($(W_{\mu}^0)^L$, $(W_{\mu}^0)^R$,
$\frac{\sqrt{2}}{2}B_{\mu}^Z$, $A_{\mu}^X$) are given by
\small
\begin{eqnarray}
\left( \begin{array}{cccc}
\frac{1}{2}g_4^2 V_1^2 &-\frac{1}{2}g_4^2V_1^2 &g_4^2V_2^2 &0 \\
-\frac{1}{2}g_4^2 V_1^2&\frac{1}{2}g_4^2V_1^2 +g_4^2 v_S^2& -g_4^2(V_2^2+v_S^2)&-g_4g_Xv_S^2 \\
g_4^2V_2^2 &-g_4^2(V_2^2+v_S^2)&2 g_4^2V_1^2 +g_4^2v_S^2& g_4g_Xv_S^2\\
0&-g_4g_Xv_S^2& g_4g_Xv_S^2& g_X^2v_S^2 \\
\end{array} \right)~,~\,
\end{eqnarray}
\normalsize
with the constants $V_1^2= v_1^2+v_2^2+v_3^2+v_4^2$ and $V_2^2= v_1^2+v_2^2-v_3^2-v_4^2$.
It can be seen from the mass matrix that the determinant vanishes,
which indicate the existence of massless gauge boson corresponding to the photon.
The eigenvector corresponding to the photon can be written as
\begin{eqnarray}
A_{\mu}=\sqrt{ 2 \left(\frac{g_X}{ g_4}\right)^2+1}
    \left\{\frac{ g_X}{g_4}[(W_{\mu}^0)^L+(W_{\mu}^0)^R]+A_{\mu}^X\right\} \, .
\end{eqnarray}
Using the expression $\sin^2\theta_w=g_X^2/(2 g_X^2+g_4^2)$ we know that
\begin{eqnarray}
A_{\mu}=\sin\theta_w[(W_{\mu}^0)^L+(W_{\mu}^0)^R]+\sqrt{\cos2\theta_w}A_{\mu}^X \, .
\end{eqnarray}
The eigenvectors and eigenvalues for other neutral gauge bosons
are rather complicated, which can be obtained numerically. We
parameterize the general mixing between neutral gauge bosons by a
unitary matrix $K_{ij}$ and the mass eigenstates are obtained by
the rotation
\begin{eqnarray}
\left( \begin{array}{c}
A_{\mu}\\ Z_{\mu}\\ Z_{\mu}^{1\prime}\\ Z_{\mu}^{2\prime} \end{array}\right)
=\left( \begin{array}{cccc}
\sin\theta_w&\sin\theta_w&0&\sqrt{\cos2\theta_w}\\
K_{21}&K_{22}&K_{23}&K_{24}\\ K_{31}&K_{32}&K_{33}&K_{34}\\
K_{41}&K_{42}&K_{43}&K_{44} \end{array}\right)
\left(\begin{array}{c}
(W^{0}_{\mu})_L\\ (W^0_{\mu})_R\\ \frac{\sqrt{2}}{2}B_{\mu}\\ A_{\mu}^X
\end{array}\right) \, .
\end{eqnarray}
From the VEV of the $\Sigma$ field, we can get the mass term for
fermions:
\small
\begin{eqnarray}
{\cal{L}}&=&\sum\limits_{ab}\left\{
y_{ab}^1\left[(U_{L})^T_bC(v_1-iv_2)U_{La}^c
      +(D_L)^T_bC(v_3-iv_4)D_{La}^c \right. \right.\nonumber \\
&& \left.+(u_L^c)^T_aC(v_1+iv_2)u_{Lb}+(d_L^c)_a^TC(v_3+iv_4)d_{Lb}\right]\nonumber \\
&& +y^2_{ab}\left[(N_L)_b^TC(v_1-iv_2)N_{La}^c+(E_L)^T_bC(v_3-iv_4)E_{La}^c\right.\nonumber \\
&& \left.+(\nu_L^c)_a^TC(v_1+iv_2)\nu_{Lb}+(e_L^c)_a^TC(v_3+iv_4)e_{Lb}\right]\nonumber\\
&&+y^1_{ab}\left[U_{La}^TCv_5u_{Lb}-D_{La}^TCv_6d_{Lb}
  -(u_L^c)_a^TCv_5\(U_{L}^c\)_b+(d_L^c)_a^TCv_6\(D_L^c\)_b\right]\nonumber\\
&&+y^2_{ab}\left[N_{La}^TCv_5\nu_{Lb}-E_{La}^TCv_6e_{Lb}
  -(\nu_L^c)_a^TCv_5\(N^c_L\)_b+(e_L^c)_a^TCv_6\(E_L^c\)_b\right]\nonumber\\
&&+y^3_{ab}\left[U_{La}^TCv_0u_{Lb}-D_{La}^TCv_0d_{Lb}
  +(u_L^c)^T_aCv_0(U_L^c)_b-(d_L^c)_a^TCv_0\(D_L^c\)_b\right]\nonumber\\
&&\left.+y^4_{ab}\left[N_{La}^TCv_0\nu_{Lb}-E_{La}^TCv_0e_{Lb}
+(\nu_L^c)_a^TCv_0(N^c_L)_b-(e_L^c)_a^TCv_0(E_L^c)_b\right]\right\}~.~\,
\end{eqnarray}
\normalsize
In principle, the flavor mixing can be obtained through the diagonalization
of the $6\times 6$ mass matrix. However, we know from the Yukawa
interaction in the Lagrangian that the flavor structure can be
decomposed as the direct product of the mixings between different
generations and the mixings within each generation.

We first analyze the mixings within each generation. There are
mixing between the standard model fermions and the mirror
fermions. We can get the mass matrix for fermions before mixing in
different generations
\begin{eqnarray} \label{ckmmatrix}
&& \left((u_L^c)_a ~~U_{La}\right)
   \left( \begin{array}{cc}
y^1_{ab}(v_1+i v_2)&y^3_{ab}v_0-y^1_{ab}v_5 \\
y^3_{ab}v_0+y^1_{ab}v_5&~~~ y^1_{ab}(v_1-iv_2) \end{array} \right)
\left( \begin{array}{c} u_{Lb}\\ (U_L^c)_b \end{array} \right) \\
&& \left((d_L^c)_a ~~D_{La}\right)
 \left( \begin{array}{cc}
 y^1_{ab}(v_3+i v_4)&-y^3_{ab}v_0+y^1_{ab}v_6 \\
-y^3_{ab}v_0-y^1_{ab}v_6& y^1_{ab}(v_3-iv_4)
 \end{array} \right)
 \left( \begin{array}{c} d_{Lb} \\(D_L^c)_b \end{array} \right)~.~\,
\label{ckmmatrix2}
\end{eqnarray}
The gauge eigenstates can be diagonalized into mass eigenstates
by bi-unitary transformations up to arbitrary phases
\begin{eqnarray}
U{\cal M}V^{\dagger}={\cal M}_d~,~\,
\end{eqnarray}
with ${\cal M}$ denoting the mass matrix in Eqs.(\ref{ckmmatrix}) and
(\ref{ckmmatrix2}).
The mixing angles in  $U$ and $V$ can be determined, e.g., for
the mass matrix in Eq.(\ref{ckmmatrix}), the two mixing angles are
given by
\begin{eqnarray}
\tan2\phi_1=-\frac{\sqrt{v_1^2+v_2^2}}{v_5} \, ,
~~~~\tan2\phi_2=\frac{\sqrt{v_1^2+v_2^2}}{v_5}\, .
\end{eqnarray}
Similar results can be obtained for the down type quarks. From the
expressions we can see that the mixing angle is independent of
$v_0$. The rotation matrices between gauge eigenstates and mass
eigenstates before CKM mixing are denoted by $U$ for $(u_{La},
U_{La}^c)$, $V$ for $(u_{La}^c, U_{La})$, $P$ for
$(d_{La},D_{La}^c)$ and $Q$ for $(d_{La}^c, D_{La})$. So we get the
couplings between one-generation fermions and the $SU(2)_L$ and
$SU(2)_R$ gauge bosons, which in the interaction states are given by
\small
\begin{eqnarray} \hspace*{-0.5cm}
{\cal L}&=&\frac{g_4}{2}\left(\bar u_L\gamma^{\mu}d_LW_{L\mu}^{+}
  -\bar U_L^c \gamma^{\mu}D_L^cW_{R\mu}^{+}-\bar U_L \gamma^{\mu}D_LW_{L\mu}^{+}
  +\bar u_L^c \gamma^{\mu}d_L^cW_{R\mu}^{+}\right)+h.c.~,~\,
\end{eqnarray}
and in the mass eigenstates are given by
\begin{eqnarray}
{\cal L} &=&\frac{g_4}{2}\left\{(W_{1\mu}^+\cos\zeta-W_{2\mu}\sin\zeta)
 \left[ U_{11}\left( P^{\dagger}_{11}\bar{u}_L\gamma^{\mu}d_L
           +P^{\dagger}_{12}\bar{u}_L\gamma^{\mu}D_L^c\right) \right.\right.\nonumber \\
&&\left. + U_{21} \left( P^{\dagger}_{11}\bar U_L^c \gamma^{\mu}d_L
      +P^{\dagger}_{12}\bar U_L^c \gamma^{\mu}D_L^c\right)
+ V_{12}\left( Q^{\dagger}_{21}\bar u^c_L \gamma^{\mu} d_L^c
     \right.\right.\nonumber \\
&&\left.\left. +Q^{\dagger}_{22}\bar u^c_L \gamma^{\mu} D_L \right)
  +V_{22}\left( Q^{\dagger}_{21}\bar U_L \gamma^{\mu} d_L^c
      +Q^{\dagger}_{22}\bar U_L \gamma^{\mu} D_L \right)\right]
\nonumber \\
&& +(W_{1\mu}^+\sin\zeta-W_{2\mu}\cos\zeta)
  \left[U_{12} \left( P^{\dagger}_{21}\bar{u}_L\gamma^{\mu}d_L
      +P^{\dagger}_{22}\bar{u}_L\gamma^{\mu}D_L^c\right)
\right.\nonumber \\
&& +U_{22}\left( P^{\dagger}_{21}\bar U_L^c \gamma^{\mu}d_L
      +P^{\dagger}_{22}\bar U_L^c \gamma^{\mu}D_L^c \right)
+V_{11}\left( Q^{\dagger}_{11}\bar u^c_L \gamma^{\mu}d_L^c \right.
  \nonumber \\
&&\left.\left.\left. +Q^{\dagger}_{12}\bar u^c_L \gamma^{\mu}D_L
\right) +V_{21}\left( Q^{\dagger}_{11}\bar U_L \gamma^{\mu} d_L^c
      +Q^{\dagger}_{12} U_L \gamma^{\mu} D_L\right) \right]\right\}~.~\,
\end{eqnarray}
\normalsize
Here it can be seen that there are additional mixing coefficients
besides CKM mixing matrix between different generations.
It is a unique feature of this model in contrast to the minimal left-right model.
To be consistent with various experiments, we require the mixing angle
is small which indicates that $v_1,v_2{\ll}v_5$ and $v_3,v_4{\ll}v_6$.
The neutral currents are also interesting due to the mixing between
the SM fermions and the new mirror fermions.

For vector type couplings, the couplings of the mass
eigenstates take the same form as the interaction states due to the
unitarity of the mixing matrix, which are given by
\begin{eqnarray}
{\cal L}_{NC}^{\rm vector}&=&\frac{2}{3} e
\left[\bar{u_L}\gamma_{\mu}u_LA_{\mu}+\bar{U_L^c}\gamma_{\mu}U_L^cA_{\mu}
-\bar{U_L}\gamma_{\mu}U_LA_{\mu}-\bar{u_L^c}\gamma_{\mu}u_L^cA_{\mu}
\right]~.~\,
\end{eqnarray}
For the chiral couplings, due to the different mixing between
$W_{L\mu}^0$ and $W_{R\mu}^0$, the couplings are different
for the mass eigenstate and the interaction states,
which in the interaction states are given by
\small
\begin{eqnarray}
{\cal L}_{NC}^{\rm chiral}=\frac{g_4}{2}\left[
\bar{u_L}\gamma_{\mu}u_LW_{L{\mu}}^0+\bar{U_L^c}\gamma_{\mu}U_L^cW_{R\mu}^0
+\bar{u_L^c}\gamma_{\mu}u_L^cW_{R{\mu}}^0+\bar{U_L}\gamma_{\mu}U_LW_{L\mu}^0
\right]~,~\,
\end{eqnarray}
\normalsize
and in the mass eigenstates are given by
\small
\begin{eqnarray}
{\cal L}_{NC}^{\rm chiral}&=&\frac{g_4}{2}
\left[  U_{11} U^{\dagger}_{11}\bar u_L\gamma_{\mu}u_L
        +U_{21}U^{\dagger}_{11} \bar U_L^c \gamma_{\mu}u_L
        +U_{11}U^{\dagger}_{12}\bar u_L \gamma_{\mu} U_L^c
 \right. \nonumber \\
&& +U_{21}U^{\dagger}_{12}\bar U_L^c\gamma_{\mu}U_L^c
        +V_{12}V^{\dagger}_{21}\bar u_L^c\gamma_{\mu}u_L^c
        +V_{22}V^{\dagger}_{21}\bar U_L \gamma_{\mu} u_L^c \nonumber \\
&& \left.+V_{12}V^{\dagger}_{22}\bar u_L^c \gamma_{\mu}U_L
        +V_{22}V^{\dagger}_{22}\bar U_L \gamma_{\mu}U_L
 \right]  K_{1j}^{\dagger}Z^j_{\mu}
 +\cdots~.~\,
\end{eqnarray}
\normalsize From $K_{11}^{\dagger}=K_{21}^{\dagger}=\sin\theta_w$
we see that the couplings of the photon can be obtained correctly.
From the Lagrangian we can see that the flavor-changing neutral
current (FCNC) is non-vanishing.

To simplify the discussion on the fermion masses and mixings, we take two special
limits to illustrate the results:
\begin{itemize}
\item[(1)] $v_0=v_5=v_6=0$:
In this case, the standard model fermions have degenerate masses with the new
mirror fermions and the mass matrix is diagonal which corresponds to
no mixing between the fermions. The degenerate mass parameter is generic
in grand unification models. We know that such tree-level relations are
hold at the unification scale. We may anticipate that the degeneracy of fermion
masses is spoiled by the renormalization group running.
From the renormalization group equation of the mass parameter (with
only contributions from the gauge parts taken into account)
 \begin{eqnarray}
\frac{d\ln{m(\mu)}}{d\ln\mu}=\sum\limits_{i} b_m^{i}g_i^2(\mu)\, ,
 \end{eqnarray}
where
 \begin{eqnarray}
b_m^{i}=-\frac{3}{8\pi^2}\sum\limits_{a}(T^aT^a)_{jk}~,~\,
 \end{eqnarray}
with $T^a$ being the representation matrices appropriate for
the fermions \footnote {For $SU(N)$ with $N{\geq}2$ we have
$\sum\limits_{a}(T^aT^a)_{jk}=\frac{N^2-1}{2 N}\delta_{jk}$;
while for $U(1)_Y$, $(T^0)^2=c^2\left(\frac{Y}{2}\right)^2$},
we can get the relation
 \begin{eqnarray}
\frac{m(\mu)}{m(\mu_0)}=
\prod\limits_{i}\left(\frac{g_i(\mu)}{g_i(\mu_0)}\right)^{-\frac{b_m^i}{b_i}}\, .
 \end{eqnarray}
Here $b_i$ is the beta function for the gauge coupling. We know that
the $SU(2)$ gauge bosons do not contribute to the fermion mass
because the right-handed (left-handed) fermions are gauge singlet
under $SU(2)_L$ ($SU(2)_R$). Then the mass degeneracy of $u,U^c$ is
not spoiled even when the RG running is taken into account. So this
scenario is not acceptable to describe our world.
\item[(2)]  $v_1=v_2=v_3=v_4=0$:
   In this case, the standard model fermions acquire different masses
in contrast to new mirror fermions. The mass matrix in the fermion
mass term $(u_L,U_L^c) {\cal M} (u_L^c,U_L)^T$ is given by
\begin{eqnarray}
{\cal M}= \left( \begin{array}{cc}
0&y_3v_0-y_1v_5 \\ y_3v_0+y_1v_5&0 \end{array} \right) \, ,
\end{eqnarray}
which is diagonalized by the rotations $(u_L,U_L^c)^T\to U
(u_L,U_L^c)^T$ and $(u_L^c,U_L)^T \to V (u_L^c,U_L)^T$ with $U$ and
$V$ given by
\begin{eqnarray}
U=\left( \begin{array}{cc}
~~1~~& 0 \\ 0 &~~1~~ \end{array} \right)\, , ~~~
V=\left( \begin{array}{cc} 0&~~1~~ \\ ~~1~~&0 \end{array}\right)~.~\,
\end{eqnarray}
Then we can get mass eigenstates with masses
\begin{eqnarray}
m_1^2=(y_3v_0-y_1v_5)^2\, ,
~~m_2^2=(y_3v_0+y_1v_5)^2 \, .
\end{eqnarray}
However, in this case the mass eigenstates of the fermions are of
vector type (for example, the $SU(2)_L$ is vector-like)
instead of chiral type and thus it cannot explain our world
with chiral fermions.
\end{itemize}
Going beyond these special limits, the most general parameters
can give non-degeneracy masses for the standard model fermions
and the new mirror fermions while at the same time the theory is
of chiral type.

Our previous discussions on the flavor structure concentrate mainly
on the mixing within each generation. As mentioned earlier, the
flavor structure can be decomposed as the direct product of the
mixings between different generations and the mixings within each
generations. The low energy CKM mixing can be obtained by
diagonalize the mass matrix ${\cal M}^u_{ab}$ and ${\cal M}^d_{ab}$.
We can see from equation ($\ref{ckmmatrix}$) that the presence of
the VEV of the singlet Higgs $S$ is necessary. Otherwise, if
$v_0=0$, there will be no CKM type mixing in the charged currents.
Note that from the coupling of the fermions with charged gauge
bosons we see that the low energy CKM matrix is not unitary.

An interesting possibility occurs when the standard model fermions
are massless at tree level. The standard model fermions can obtain
masses through loops involving heavy mirror fermions and Higgs fields
etc. Therefore, we anticipate that the standard model fermions acquire
much smaller masses than heavy mirror fermions. The Yukawa
couplings can also induce the quark mixings and CP violation between
different generations. The spontaneous breaking of the symmetric
tensor Higgs field $\Delta$ in representation (${\bf 10, 2}$) of
$SU(4)_W{\times}U(1)_{B-L}$ can also give Majorana mass terms for
right-handed neutrinos. The additional new $SU(4)_W{\times}U(1)_{B-L}$
invariant Yukawa coupling terms are given by
\begin{eqnarray}
{\cal L}_{Yukawa}=\sum\limits_{f}y_3^f(N_f)^T_iC\Delta^{ij}(N_f)_j+h.c.~.~\,
\end{eqnarray}
The VEV of $\Delta$ gives Majorana mass terms for the right-handed
neutrinos. So we can get light Majorona neutrino masses after diagonalizing
the mass matrix. In case of the tree-level massless standard model
fermions, it is natural that the loop-induced Dirac mass of the neutrinos
is of the same order as $m_e$. We can estimate that
\begin{eqnarray}
\left( \begin{array}{cc} 0&m_D\\ m_D&m_N\\ \end{array} \right)
& {\Longrightarrow}& m_{\nu}{\sim}\frac{m_D^2}{m_N}{\sim}
                     \frac{m_e^2}{m_N}{\sim}10^{-1} {\rm ~eV} ~,~\,\nonumber \\
&{\Longrightarrow}& m_N=y_3^fv_S {\sim}{\cal O}(1)  {\rm ~TeV} \, .
\end{eqnarray}
We know that the VEV of $\Delta$ also breaks the $SU(2)_R$ symmetry.
From the Majorana mass scale, we can get that the typical mass scale
of the $SU(2)_R$ gauge boson $M_{W_R}$ to be higher than several TeV.
In general cases, the mass for $M_{W_R}$ can be much heavier.

It is well known that in Pati-Salam model leptons can be seen as the
fourth color so that $U(1)_{B-L}$ and $SU(3)_C$ can be unified into
$SU(4)_{PS}$ gauge group. Depending on the different symmetry breaking
scales, the gauge symmetry breaking patterns have the following
possibilities:
\begin{itemize}
\item[(a)] One possibility is
\begin{eqnarray*}
SU(4)_{PS}{\times}SU(4)_W&\rightarrow&
SU(3)_C{\times}U(1)_{B-L}{\times}SU(4)_W\\&\rightarrow&
SU(3)_C{\times}U(1)_{B-L}{\times}SU(2)_L{\times}SU(2)_R{\times}U(1)_Z\\
&{\rightarrow}& SU(3)_C{\times}U(1)_Q~.~\,
\end{eqnarray*}
This symmetry breaking pattern is just what we have discussed.
\item[(b)] The other possibility is
\begin{eqnarray*}
SU(4)_{PS}{\times}SU(4)_W&{\rightarrow}&
SU(4)_{PS}{\times}SU(2)_L{\times}SU(2)_R{\times}U(1)_Z
\\&{\rightarrow}&
SU(3)_C{\times}U(1)_{B-L}{\times}SU(2)_L{\times}SU(2)_R{\times}U(1)_Z\\
&{\rightarrow}& SU(3)_C{\times}U(1)_Q~.~\,
\end{eqnarray*}
 This case is interesting because the intermediate steps contain
the Pati-Salam model. It can induce new type of unification besides
$SO(10)$. Also the representation of the matter fields can be
economically written as $X,Y$ in representations $\bf{(4,4)}$
 and $\bf{(4,4^*)}$ of $SU(4)_{PS}{\times}SU(4)_W$ respectively
because the leptons can be regarded as the fourth color of quarks.
In the Pati-Salam model the gauge coupling $g_{2L}=g_{2R}$ is fixed
by a discrete symmetry which otherwise holds only in the unification
scale when the gauge groups fit into $SO(10)$.
In our $SU(4)_W{\times}U(1)_{B-L}$ unification theory, such identical gauge
strength is the consequence of the relatively low energy gauge
unification.
\end{itemize}
Our model is anomaly free. We can check that different kinds of
triangle anomalies cancel:
\begin{itemize}
\item For $SU(4)_W-SU(4)_{w}-SU(4)_W$ we have
\begin{eqnarray}
Tr(T_a\{T_b,T_c\})&=&\frac{1}{2}A(R)d_{abc} \, ,\\
A(4)+A(4^*)&=&0 \, ,
\end{eqnarray}
where $d_{abc}$ is the totally symmetric tensor in the
anticommutators of fundamental representation
\begin{eqnarray}
\{\lambda_a,\lambda_b\}=2 d_{abc}\lambda_c~.~\,
\end{eqnarray}
\item For $U(1)_{B-L}-SU(4)_{w}-SU(4)_W$ we have
\begin{eqnarray}
 \sum\limits_{fermion}Y_{B-L}=0 \, .
\end{eqnarray}
\item For $U(1)_{B-L}-U(1)_{B-L}-U(1)_{B-L}$ we have
\begin{eqnarray}
 \sum\limits_{fermion}Y_{B-L}^3=0 \, .
\end{eqnarray}
\end{itemize}
If we assume further unification of $SU(4)_{PS} \times SU(4)_W$ at scale $M_U$,
we can give prediction for $\sin^2\theta_w$.
 From the normalization conditions we know that
\begin{eqnarray}
 g^2_{B-L}=\frac{3}{2}g_4^2~,~\,
\end{eqnarray}
holds at the $SU(4)_{PS}$ unification scale.
Then we can predict
\begin{eqnarray}
\sin^2\theta_w=\frac{g_{B-L}^2}{g_L^2+2g_{B-L}^2}=\frac{3}{8}~,~\,
\end{eqnarray}
which holds at $\mu=M_U$. It is interesting to note that this value is
same as the $SU(5)$ unification prediction. The prediction of
$\sin^2\theta_w$ at weak scale depends on the symmetry breaking chains.
We know that the running of the gauge couplings is
\begin{eqnarray}
\frac{1}{\alpha_i(\mu^2)}=\frac{1}{\alpha_i(M^2)}-\frac{b_i}{4\pi}\ln\frac{\mu^2}{M^2}\, ,
\end{eqnarray}
where
\begin{eqnarray}
b_i=-(\frac{11}{3}C_2(G)-\frac{4}{3}\sum\limits_{r^{\prime}}^{n_f}C(r^{\prime})
   -\sum\limits_{r}^{n_h}\frac{1}{3}C(r))~,~\,
\end{eqnarray}
with $n_f$ being the number of fermion flavors, $n_h$ being the number of
complex scalar fields, and $C_2(G)=N$ being the quadratic Casimir operator
of the adjoint representation of $SU(N)$. The Casimir $C(r)$ is defined by
\begin{eqnarray}
Tr[T_r^aT_r^b]=C(r)\delta^{ab}~,~\,
\end{eqnarray}
with $C(G)=N$ in the adjoint representation;
$C([2])=(N+2)/2$ and $C([1^2])=(N-2)/2$ for symmetric and
antisymmetric representations, respectively.

The key difference between the runnings of the two $SU(4)$ gauge couplings
lies in the Higgs contributions. We can introduce the adjoint representation
Higgs of $SU(4)_{PS}$ to break such gauge symmetry into
$SU(3)_C \times U(1)_{B-L}$. Similarly, in our case, the $SU(4)_W$
gauge symmetry is broken by the adjoint Higgs field $\Phi$. We can
extend the Higgs field $\Delta$ as representation $\bf{(10,10)}$ of
$SU(4)_{PS} \times SU(4)_W$. The matter field $X,Y$ lie in
$\bf{(4,4)}$ and $\bf{(4,4^*)}$ representations respectively which
give identical contributions to both $SU(4)$ gauge coupling runnings.
Suppose the couplings are unified at the mass scale $M_U$, from
\begin{eqnarray}
\frac{d(\alpha_4^{w})}{d\ln\mu}&=&b_{4}^{w}\frac{(\alpha_4^{w})^2}{2\pi}\, ,\\
\frac{d(\alpha_4^{PS})}{d\ln\mu}&=&b_4^{PS}\frac{(\alpha_4^{PS})^2}{2\pi}\, ,
\end{eqnarray}
we obtain
\begin{eqnarray}
\frac{1}{\alpha_4^{PS}(\mu^2)}-\frac{1}{\alpha_i^{w}(\mu^2)}
=-\frac{b_4^{PS}-b_4^{w}}{2\pi}\ln\frac{\mu}{M_U}
=\frac{2}{3\pi}\ln\frac{\mu}{M_U}~.~\,
\end{eqnarray}
Given the symmetry breaking pattern, we can predict $\sin\theta_w$
at weak scale through renormalization group running.

\section{Renormalization group running of gauge couplings}
\label{sec-3}
  We now discuss the renormalization group running
  of the gauge couplings in different scenarios, including
  the orbifold breaking cases and the pure four-dimensional
  model with the Higgs mechanism.
   We use the inputs \cite{PDG}
 \beqa
  M_Z&=&91.1876\pm0.0021 ~,~\,\\
   \sin^2\theta_w(M_Z)&=&0.2312\pm 0.0002 ~,~\,\\
  \alpha^{-1}_{em}(M_Z)&=&127.906\pm 0.019 ~,~\,\\
  \alpha_3(M_z)&=&0.1187\pm 0.0020 \, .
  \eeqa
From the electroweak theory we get the couplings at scale $M_Z$
\beqa
\alpha_1(M_Z)&=&\f{\alpha_{em}(M_Z)}{\cos^2\theta_w} ~,~\,\\
\alpha_2(M_Z)&=&\f{\alpha_{em}(M_Z)}{\sin^2\theta_w}~,~\,\\
\alpha_s(M_Z)&=&\f{g_s^2}{4\pi} ~.~\,\eeqa
The renormalization group running of the gauge couplings reads:
\beqa
\f{d~\alpha_i}{dt}=\f{b_i}{2\pi}\alpha_i^2 \, .
\eeqa
At the scale of the $SU(2)_R$ gauge boson mass $M_R$, the left-right
  $SU(3)_C\times SU(2)_L\times SU(2)_R\times U(1)_{B-L}$ symmetry breaks
into the standard model gauge groups. From the symmetry breaking
chain and the kinematic terms we write, we know the relation \beqa
\f{1}{e^2}=\f{1}{g_{2L}^2}+\f{1}{g_{2R}^2}+\f{1}{g_{B-L}^2} \, .
\eeqa Then we can get the coupling $g_{B-L}$ at the scale $M_R$.

\subsection{Beta functions in five-dimensional orbifold}
We calculate the beta functions for $SU(2)_R$ ($g_L=g_R$ at scale
$M_R$) in orbifold breaking scenario. The difference of matter
spectrum
 between inner and outer automorphism orbifold breakings lies in the
 fact that there is no $U(1)_Z$ in outer automorphism orbifold
 breaking scenario. As we know, the left-right symmetry guarantees
 that the two gauge couplings are the same at the energy scale
 $M_R$. So in the outer-automorphism orbifold breaking case, the
 unification scale, which depends only on the compactification scale $1/R$,
 can be much lower than in the inner automorphism orbifold
 breaking. Here we only discuss  the gauge coupling unification
 in the inner automorphism case.

\begin{figure}[htb]
\begin{center}
\scalebox{0.7}{\epsfig{file=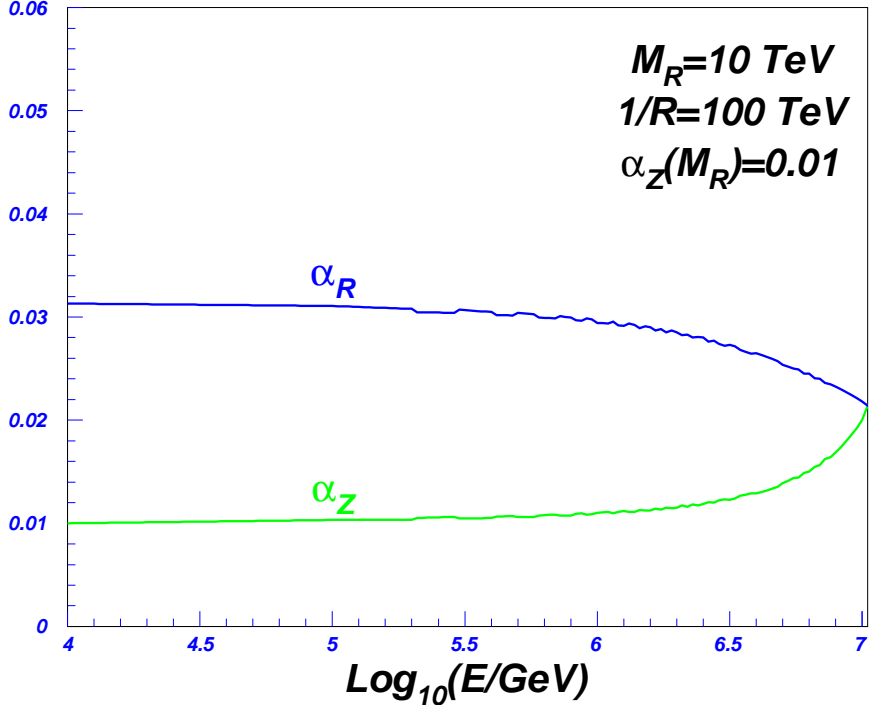}}
\end{center}
\vspace*{-1.0cm}
\caption{\small The gauge coupling unification in the inner automorphism
 orbifold breaking case. The lower and upper curves are the running of
the couplings of $U(1)_Z$ and $SU(2)_R$, respectively. }
\label{RGE}
\end{figure}

 The $SU(2)_R$ coupling at scale $E$ is given by
 \beqa
 \alpha^{-1}(E)&=&\alpha^{-1}(M_R)-\f{1}{2\pi}\f{2}{3}\ln\(\f{M_R}{E}\)
 -\f{1}{2\pi}\f{2}{3}\sum\limits_{n=1}^{k}\ln\(\f{2 n }{E R}\)\Theta(E-\f{2n}{R}) \nonumber \\
&&-\f{1}{2\pi}2\sum\limits_{n=0}^{k}\ln\(\f{2 n+1 }{E R}\)\Theta(E-\f{2n+1}{R}) \, ,
 \eeqa
 where $\Theta(x)$ is the step function defined as
 \beqa
 \Theta(x)=\left\{ \begin{array}{c} 1~~~x\ge0\\ 0~~~x<0 \end{array} \right .~.~\,
 \eeqa
 The $U(1)_Z$ coupling at scale $E$ is
 \beqa
 \alpha^{-1}(E)&=&\alpha^{-1}(M_R)+\f{1}{2\pi}\f{26}{3}\ln\(\f{M_R}{E}\)
 +\f{1}{2\pi}\f{26}{3}\sum\limits_{n=1}^{k}\ln\(\f{2 n }{E R}\)\Theta(E-\f{2n}{R})\nonumber\\
&&-\f{1}{2\pi}6\sum\limits_{n=0}^{k}\ln\(\f{2 n+1 }{E R}\)\Theta(E-\f{2n+1}{R})\, .
 \eeqa
We anticipate that the unification of the two gauge couplings occurs not
too high above the compactification scale $1/R$
due to the fact that more and more KK modes give contributions.
An example is shown in Fig. \ref{RGE} for fixed values of $M_R$ and $\alpha_Z(M_R)$.

\subsection{Unification of $SU(4)_{PS}$ with $SU(4)_W$}
We now discuss the unification of the four-color $SU(4)_{PS}$ with
$SU(4)_W$ in four dimensions.
When $U(1)_{B-L}$ is embedded into the $SU(4)_{PS}$ group, the charge
should be normalized according to $SU(4)$ generator as
\begin{eqnarray}
 \f{1}{2}Y_{B-L}=\left(\begin{array}{cccc}
\f{1}{6}&&&\\ &\f{1}{6}&&\\ &&\f{1}{6}&\\ &&&-\f{1}{2} \end{array}\right)
=\sqrt{\f{2}{3}}T_{B-L}\, .
\end{eqnarray}
From the relation
\beqa g^{PS}_4 T_{B-L}=g_{B-L}\f{1}{2}Y_{B-L}~,~\,
\eeqa
we get the normalization factor
\beqa
g_{B-L}=\f{\sqrt{6}}{2}g_{4} \, .
\eeqa
We know that at low energy $SU(4)_{PS}$ is broken into $SU(3)_C\times U(1)_{B-L}$.
Here we assume that this step is accomplished through the VEV of the $SU(4)_{PS}$
adjoint Higgs field $\Sigma_2$. They naturally acquire masses of order of the
breaking scale. Besides, we know from previous discussions that the symmetric
Higgs field $\Delta$, which carries the $U(1)_{B-L}$ charge $2$, is used to break
the $SU(2)_R\times U(1)_{B-L}$. So to unify $U(1)_{B-L}$ with $SU(3)_C$ into
$SU(4)_{PS}$, the symmetric Higgs $\Delta$ must be extended to the $SU(4)_{PS}$
representation. According to the two $SU(4)$, the matter content is
$X\({\bf 3,4}\)_{\f{1}{3}} \oplus L\({\bf 1,4}\)_{-1}=({\bf 4,4})$ and
$Y\({\bf 3,\bar{4}}\)_{-\f{1}{3}} \oplus N\({\bf 1,\bar{4}}\)_{1}=({\bf 4,\bar{4}})$;
while the scalar parts given by $\Phi({\bf 1,15})$, $\Delta({\bf  1,10})$ extend to
$({\bf 10,10})$ and $\Sigma({\bf 1,15})$.
There are two possibilities for the symmetry breaking chain:
\begin{itemize}
\item[(1)] In most cases, we are interested in the case that the partial
unification $SU(4)_W$ at scale $M_{U}^W$ is not too high. Interestingly,
 the four-color $SU(4)_{PS}$ breaking scale $M_U^{PS}$ can also be
around $M_{U}^w$ because the $SU(4)_{PS}$ gauge bosons will not
generate the proton decay. After $SU(4)_{PS}$ is broken down to
$SU(3)_C\times U(1)_{B-L}$, the symmetric Higgs $({\bf 10,10})$ is
decomposed into the representations in $SU(3)_C$ and $SU(4)_W$):
$({\bf 10,10})=({\bf 6,10})_{\f{2}{3}}\oplus({\bf 1,10})_{-2}
\oplus({\bf 3,10})_{-\f{2}{3}}$. We know from the cross mixing
terms in the most general Higgs potential that these Higgs fields
acquire masses of order $M_U^{PS}$ except that we fine-tune the
symmetric Higgs field $\Delta({\bf 1,10})$ to be of the order of
the right handed gauge boson mass $M_R$. We can also possibly tune
the masses of the Higgs fields $({\bf 3,10})_{-\f{2}{3}}$ and
$({\bf 6,10})_{\f{2}{3}}$ to be of the order $M_U^w$. Then only
the adjoint Higgs $\Sigma_2$ masses are at order $M_U^{PS}$.
\item[(2)] The other possibility is to decompose the symmetric
Higgs according to $SU(2)_L\times SU(2)_R\times U(1)_Z$ as $({\bf
10,10})=({\bf 10,3,1})_{1}\oplus({\bf 10,1,3})_{-1} \oplus({\bf
10,2,\bar{2}})_{0}$ if $M_U^w$ is higher than $M_U^{PS}$. Due to
the cross mixing terms, the relevant Higgs fields also acquire
masses of order $M_U^w$ scale. We can also fine tune these Higgs
masses to be of order $M_U^{PS}$ so that they may be included in
the most general Higgs potential in $SU(4)_{PS}$ symmetry broken.
\end{itemize}

In both cases, we require the mirror fermion masses at order
$M_R$. Higgs boson from $\Phi$ have masses of order $M_U^w$
and are integrated out at scale $M_R$. So at the threshold
scale $M_R$, the Higgs boson content contains
(i) the $SU(2)_L$ triplet
and $SU(2)_R$ triplet from $\Delta$,
(ii) the $SU(2)_L$ and $SU(2)_R$ adjoint representation from $\Sigma$,
and (iii) two bi-doublets $({\bf 2,2})$ under $SU(2)_L\times SU(2)_R$
from $\Sigma$ and one from $\Delta$.

In case of $M_U^w<M_U^{PS}$, the matter content at the threshold
$M_U^w$ contains the adjoint Higgs $\Sigma_1$ and $\Phi$, the
symmetric Higgs $\Delta$, and possibly $({\bf 3,10})_{-\f{2}{3}}$
and $({\bf 6,10})_{\f{2}{3}}$. If $M_U^{PS}<M_U^w$, the matter
content at the threshold $M_U^{PS}$ contains those at $M_R$ and
possibly the $({\bf 10,3,1})_{1}$,$({\bf 10,1,3})_{-1}$ and $({\bf
10,2,\bar{2}})_{0}$.

So we get the beta functions for each coupling:
\begin{itemize}
\item For $M_Z<E<M_R$, the $U(1)_Y,SU(2)_L,SU(3)_C$ beta-functions are given by
\beqa (b_1,b_2,b_3)=\(\f{41}{10},-\f{19}{6},-7\) \, . \eeqa
\item For $M_R<E<M_{U}$, the $U(1)_Z,U(1)_{B-L},SU(2)_L=SU(2)_R,SU(3)_C$ beta functions
      are given by
\beqa
(b_0,b_1,b_2,b_3)=\(\f{31}{3},13,3,-3\)\, ,
\eeqa
where $M_U={\rm min}(M_U^w,M_U^{PS})$. If the symmetry breaking chain
is $M_U^w<E<M_{U}^{PS}$, the $U(1)_{B-L},SU(3)_C,SU(4)_W$
beta-functions are
\beqa
(b_1,b_2,b_3)=\(13,-3,-3\)\, ,
\eeqa
or
\beqa
(b_1,b_2,b_3)=\(18,7,6\)~,~\,
\eeqa
if we take into account the scalar
$({\bf 3,10})_{-\f{2}{3}}$ and $({\bf 6,10})_{\f{2}{3}}$ at $M_U$.
If the symmetry breaking chain is $M_U^{PS}<E<M_{U}^{w}$, the
$U(1)_Z,SU(2)_L=SU(2)_R,SU(4)_{PS}$ beta functions are
 \beqa
(b_0,b_1,b_2)=\(\f{31}{3},3,-\f{16}{3}\)~,~\,
\eeqa
 or
\beqa
(b_0,b_1,b_2)=\(\f{77}{6},13,\f{14}{3}\)~,~\,
\eeqa if we take into account the scalar $({\bf 10,3,1})_{1}$,
$({\bf 10,1,3})_{-1}$ and $({\bf 10,2,\bar{2}})_{0}$.
\item  For $M_{U}<E$, the  $SU(4)_W$ and $SU(4)_{PS}$ beta functions are
\beqa
(b_1,b_2)=\(6,\f{14}{3}\)~,~\,
\eeqa
where $M_U^2={\rm max}(M_U^w,M_U^{PS})$.
\end{itemize}
In Table \ref{unification} we show the $SU(4)_W$ unification scale
$M_U^w$ for various values of $M_R$ and $\alpha_Z(M_R)$ in four
dimensions.
\begin{table}
\caption{$SU(4)_W$ unification scale $M_U^w$ (GeV) for various
values of $M_R$ and $\alpha_Z(M_R)$ in four dimensions. "No" means
no such a unification (the fourth color unification may occur
first). Here we assume no intermediate states between $M_U^w$ and
$M_R$, which occurs for $M_U^{w}<M^{PS}_U$.}
\begin{center}
\begin{tabular}{|c|c|c|c|c|c|}
\hline
$M_R {\Large \backslash } \alpha_Z^{-1}$ & $70.0$&$60.0$&$50.0$&$40.0$&$35.0$\\
\hline 1 TeV &$3.91\times10^{17}$&$7.41\times10^{13}$& $1.41\times 10^{10}$&$2.69\times10^6$&$3.71\times10^4$\\
\hline 5 TeV &$0.97\times10^{18}$&$1.85\times10^{14}$&$3.54\times 10^{10}$&$6.65\times10^6$&$9.31\times10^4$\\
\hline 10 TeV&$1.45\times10^{18}$&$2.76\times10^{14}$&$5.37\times10^{10}$&$0.99\times10^7$&$1.38\times10^5$\\
\hline $10^2$ TeV&$5.37\times10^{18}$&$1.02\times10^{15}$&$1.94\times10^{11}$&$3.71\times10^7$&$5.12\times10^5$\\
\hline $10^3$ TeV &$1.99\times10^{19}$&$3.91\times10^{15}$&$7.24\times10^{11}$&$1.38\times10^8$&$1.88\times10^6$\\
\hline $10^4$ TeV&$7.40\times10^{19}$&$1.40\times10^{16}$&$2.64\times10^{12}$&$5.06\times10^8$& No \\
\hline $10^5$ TeV&$2.71\times10^{20}$&$5.12\times10^{16}$&$0.98\times10^{13}$&$1.88\times10^9$& No\\
\hline
\end{tabular}
\end{center}
\label{unification}
\end{table}
\section{Phenomenology discussions}
\label{sec-4}
If we use the five-dimensional orbifold boundary conditions to
break the gauge group, the matter content at low energy scale is almost
same as in the minimal left-right model. The mirror fermions are projected
out and do not appear in the low energy effective theory.
If the space-time compactification scale $1/R$ is relatively low, different
KK modes will have various quantum corrections. However, in this case
the Landau poles will appear at low energy for some gauge couplings.
So we assume the compactification scale is not too low.
Therefore, the phenomenology in orbifold breaking scenario will be
almost the same as in the left-right model, which has already been
studied in detail in the literature.

So we concentrate on the four-dimensional case for phenomenology
discussions. In this scenario the phenomenology is also quite
similar to the minimal left-right model except that our unification
scenario at low energy predicts the mixing between the SM fermions
and the heavy mirror fermions. As discussed previously, such mixings
are independent of the generation and do not alter the CKM matrix
between different generations. However, the CKM matrix is defined
through the left-handed charged currents (strictly speaking, due to
the left-right mixing of charged gauge bosons, the CKM matrix of the
SM is defined as the coupling between lower mass eigenstates of
left-handed fermions and the lower mass $SU(2)$ charged gauge
boson), and thus is not unitary.Besides,the mirror fermions can be
heavy enough to avoid the bounds by precision tests in LEP and
Tevatron. In the following we briefly discuss some phenomenology
which is beyond the predictions of the minimal left-right models.
\begin{itemize}
\item[(1)] $K_L-K_S$ mass difference:
One of the most stringent constraints may come from the
$K_L-K_S$ mass difference. From the expression of the $SU(2)$ charged
currents we can get the leading effective operators that contribute
to $K_L-K_S$ mixing from $W_1^{\mu}$ exchange:
\begin{eqnarray}
{\cal O}_{1}&=&\bar{s_L}\gamma_{\mu}d_L \bar{s_L}\gamma_{\mu}d_L \, .
\end{eqnarray}
 From the leading order amplitudes we can get the following effective Hamilton for
$K_L-K_S$ mixing
\begin{eqnarray}
H_W^{{\Delta}S=2}=-\frac{G_F}{\sqrt{2}}\frac{\alpha}{\pi\sin^2\theta_w}
 \sum\limits_{i,j}\alpha_1^4\beta_i\beta_jC(r_i,r_j){\cal O}_1~,~\,
\end{eqnarray}
where $\beta_i=N_{id}^*N_{is}$ with $N_{ij}$ denoting the CKM mixing for left-handed fermions,
$\alpha_1$ is defined from the $SU(2)$ charged current and given by
\begin{eqnarray}
\alpha_1&=&\left[(U)_{11}(P^{\dagger})_{11}\cos\zeta+(U)_{12}(P^{\dagger})_{11}\cos\zeta\right] \, ,
\end{eqnarray}
and $C(r_i,r_j)$ is defined as
\begin{eqnarray}
C(r_i,r_j)&=&\frac{f(r_i)-f(r_j)}{r_i-r_j} ~,~\,\\
f(r_i)&=&\frac{1}{1-r_i}+\frac{r_i^2{\ln}r_i}{(1-r_i)^2}~,~\,
\end{eqnarray}
with $r_i=m_i^2/m_{W1}^2$.
Note that the effective Hamilton differs from the SM constribution by an additional
factor $\alpha_1^4$.

Due to the mixing between the new mirror fermions and the SM fermions,
there are new additional contributions illustrated in Fig. \ref{mixing}
besides those in the minimal $SU(2)_L{\times}SU(2)_R{\times}U(1)_{B-L}$
left-right model. So the experimental data of $K_L-K_S$ mixing will constrain
the masses and mixings of the mirror fermions.
\begin{figure}[htb]
\begin{center}
\epsfig{file=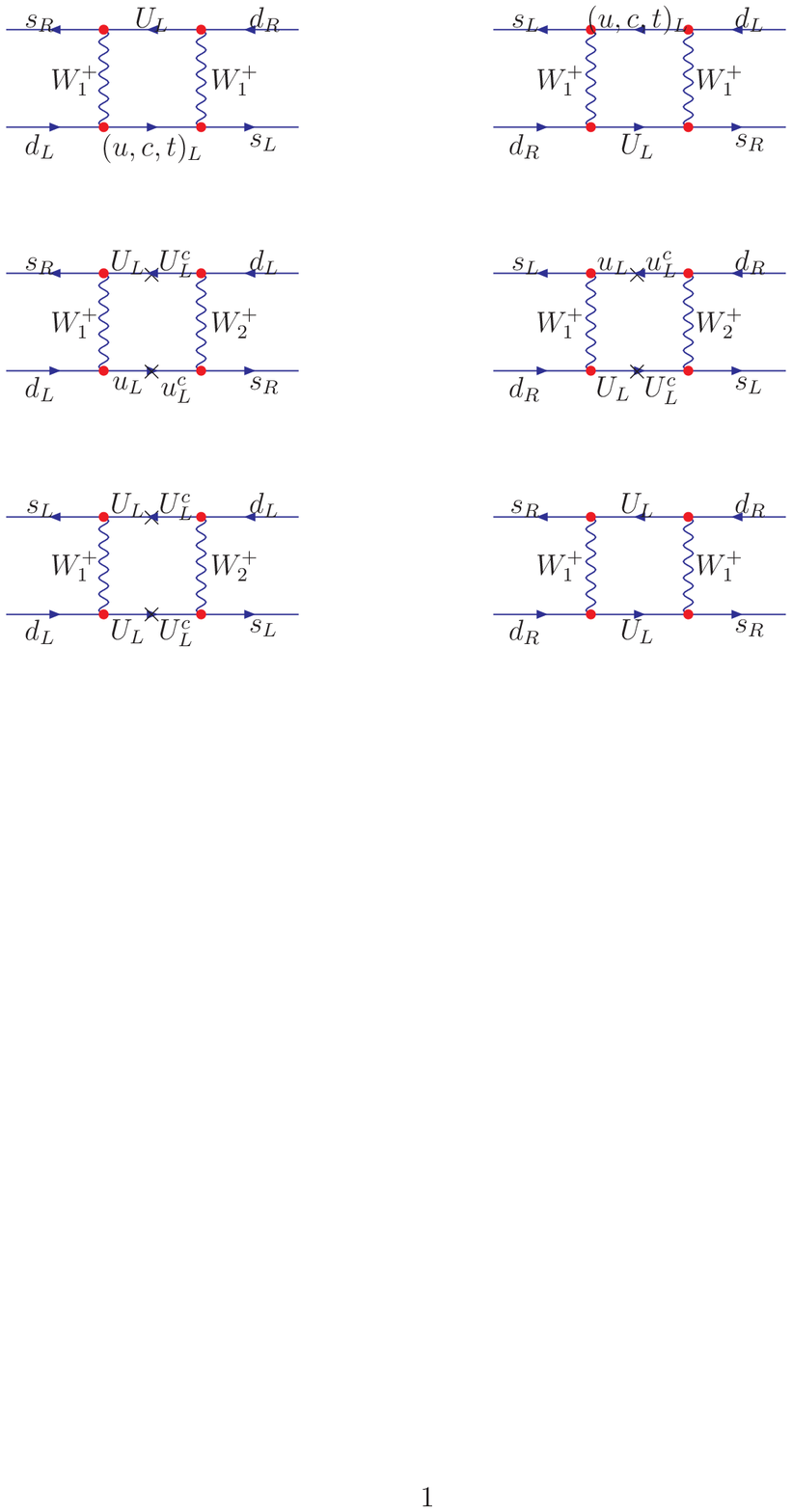, width=10cm}
\end{center}
\vspace*{-1.0cm}
\caption{ Typical additional new diagrams that contribute to
$K_L-K_S$ mixing besides that of minimal left-right model.}
\label{mixing}
\end{figure}

\item[(2)] Contributions to the neutron electric dipole moment $d_n^e$ through $W_L-W_R$
           mixing and fermion mixing \cite{edm}.
           Our model contributes to $d_n^e$ at one-loop
  level which differs greatly from the SM in which the non-vanishing
  contributions arise at three-loop level.
  The contributions also differ from the minimal
  left-right model due to the mixing between the standard fermions and
  heavy mirror fermions. So  $d_n^e$ may stringently constrain the parameter
  space of our model.

\item[(3)] Constraints from the new gauge bosons $Z^{\prime}$.
From the mass matrix of the neutral gauge bosons, we can find that
there are two additional $Z'$ gauge bosons.
The experiment constraints on $Z^{\prime}$ \cite{tosa} will constrain
the parameter space of our model.

\item[(4)] Constraints from $b{\rightarrow}s+{\gamma}$.
    The decay $b{\rightarrow}s+\gamma$ is sensitive to new physics.
    In our scenario, as discussed in the vector-type couplings,
    $b{\rightarrow}s+{\gamma}$ is still vanishing at tree-level.
    At loop level, due to the various mixings in gauge boson sector and
    the fermion sector, there are various new contributions.
    So $b{\rightarrow}s+{\gamma}$ may constrain the parameter space of our model.

\item[(5)]  Dirac neutrino and CP violation in lepton sector.
In our model it is possible for the Higgs field $\Delta$ to acquire VEVs
which also give heavy Majorana mass for the neutral components of the
new heavy fermions. In this case, through the see-saw mechanism, the
additional neutrinos can acquire light masses. So there are
totally six kinds of light neutrino species in this scenario.
It is well known that the number of neutrino species
is strictly constrained  by the Big Bang Nucleosythesis (BBN) \cite{bbn1}.
Although the non-standard BBN limit on the number of neutrino species
is relaxed to seven \cite{bbn2}, we restrain to consider the case with only
three light neutrinos, as we discussed previously. We know from
previous discussions that in lepton sector we give heavy Majorana
masses only for the standard model neutrinos and not for the new
types of neutrinos. In the heavy lepton sector we have heavy Dirac neutrinos,
which can be pair produced through $Z$ gauge boson at the LHC or ILC, and also
we have CKM-like matrix which may be measurable through some CP-violating process.

\item[(6)]  Tree-level FCNC in mirror fermion sector.
   In the sector of the heavy mirror fermions, there are tree-level FCNC
   interactions, which can induce various FCNC processes at the LHC.
   Such FCNC processes can be used to constrain the mixing angles like $\phi_i$
   which appear in the mixing  between the standard model fermions and new fermions.
\end{itemize}

\section{Conclusions}
\label{sec-5}
Left-right model is proposed to explain the parity asymmetry in
Standard Model. To understand the origin of left-right symmetry, we
studied an partial unification model based on
 $SU(4)_W{\times}U(1)_{B-L}$ which can break to the minimal left-right
 model either through the Higgs mechanism in four dimensions or through
  orbifolding in five dimensions,especially we propose to use the
 rank reducing outer automorphism orbifolding breaking mechanism.
 We scrutinized all these breaking
mechanisms and found that for the orbifold breaking in five
dimensions, the rank-reducing outer automorphism is better than the
inner automorphism and can make the low energy theory free of the
$U(1)_Z$ anomaly.It is possible for the outer automorphism
orbifolding breaking mechanism to be non-anomalous without
Chern-Simons terms and new localized fermions.For the
four-dimensional model with the Higgs mechanism, we studied in great
detail both its structure and its typical phenomenology. It turns
out that this four-dimensional scenario may predict some new
phenomenology since the new mirror fermions (which are introduced in
order to fill the SM fermions into $SU(4)_W$ without anomaly) are
preserved at low energy scale with mixing with the SM fermions. We
also examined the running and the unification of the gauge couplings
in each case, and discussed the possibility for unifying this
partial unification group with $SU(4)_{PS}$ to realize a grand
unification.

\section*{Acknowlegments}
We are grateful to Csaba Balazs, Yu-Ping Kuang and Zhong-Yuan Zhu
for enlightening discussions. This work was supported in part by the
Australian Research Council under project DP0877916, by the National
Natural Science Foundation of China under grant Nos. 10821504,
10725526 and 10635030, and by the Cambridge-Mitchell Collaboration
in Theoretical Cosmology.

\end{document}